\documentclass{aa}  

\usepackage{graphicx}
\usepackage{txfonts}

\usepackage{color}

\usepackage[authormarkup=none]{changes}
\definechangesauthor[color=red,name={RZ}]{ra}
\definechangesauthor[color=blue,name={KW}]{ko}

\usepackage{units}

\begin{document} 

\title{Mars moon ephemerides after \textcolor{black}{14} years of Mars Express data}

\author{V. Lainey
\inst{1}
\and
A. Pasewaldt
\inst{2}
\and
V. Robert
\inst{3,1}
\and
P. Rosenblatt
\inst{4}
\and 
R. Jaumann
\inst{2}
\and 
J. Oberst
\inst{2}
\and 
T. Roatsch
\inst{2}
\and
K. Willner
\inst{2}
\and
R. Ziese
\inst{5}
\and
W. Thuillot
\inst{1}
}

\institute{IMCCE, Observatoire de Paris, PSL Research University, CNRS,  Sorbonne Universit\'es, UPMC Univ. Paris 06, Univ. Lille\\
\email{valery.lainey@obspm.fr; vincent.robert@obspm.fr; william.thuillot@obspm.fr}
\and
German Aerospace Center (DLR), Rutherfordstr. 2, 12489 Berlin-Adlershof, Germany\\
\email{konrad.willner@dlr.de}
\and
IPSA, 63 bis boulevard de Brandebourg, 94200 Ivry-sur-Seine, France
\and
Laboratoire De Planétologie Et Géodynamique, Bâtiment 4, 2 Chemin de la Houssinière, 44300 Nantes, France\\
\email{Pascal.Rosenblatt@univ-nantes.fr}
\and
Technical University of Berlin, Straße des 17. Juni 135, 10623 , Berlin, Germany\\
\email{ziese@tu-berlin.de}
}

\date{Received ; accepted }

\abstract
{The Mars Express (MEX) mission has been successfully operated around Mars since 2004. Among many results, MEX has provided some of the most accurate astrometric data of the two Mars moons, Phobos and Deimos. In this work we present new ephemerides of Mars' moons benefitting from all previously published astrometric data to the most recent MEX SRC data. All in all, observations from 1877 until \textcolor{black}{2018} and including spacecraft measurements from Mariner 9 to MEX were included. Assuming a homogeneous interior, we fitted Phobos' forced libration amplitude simultaneously with the Martian tidal $k_2/Q$ ratio and the initial state of the moons. Our solution of the physical libration $1.09 \pm 0.01$ degrees deviates notably from the homogeneous solution. But considering the very low error bar, this may essentially suggest the necessity to consider higher order harmonics, with an improved rotation model, in the future. While most data could be successfully fitted, we found a disagreement between the Mars Reconnaissance Orbiter and the Mars Express astrometric data at the kilometer level probably associated with a biased phase correction. The present solution precision is expected at the level of a few hundreds of meters for Phobos and several hundreds of meters for Deimos for the coming years. The real accuracy of our new ephemerides will have to be confirmed by confrontation with independent observational means.}


\maketitle



\section{Introduction}

Space astrometry of the natural satellites allows us to reach a precision incomparable with respect to the telescopic observations. Regarding Phobos, we can access a short history of these telescopic observations in \citet{2014P&SS..102....2P}. In order to exploit the combination of space data and ground-based ones for computing ephemerides, we received funding in the scope of the European FP7 program and coordinated the European Satellite Partnership for Computing Ephemerides (ESPaCE) \citep{2012ASPC..461..659T} consortium from 2011 to 2015. ESPaCE brought together seven laboratories (IMCCE-Paris Observatory, Royal Observatory of Belgium ROB, Joint Institute for VLBI in Europe JIVE-ERIC, Technical University of Berlin TUB, German Aerospace Center DLR, French Space Center CNES, Technical University of Delft TUD), that focused on different activities related to the positioning of spacecraft and natural satellites such as digitizing of old telescopic observations, the exploitation of spacecraft radio tracking data to reconstruct spacecraft orbits and the development of planetary moons ephemerides. The ESPaCE portal at http://espace.oma.be provides access to further information. In this context, a fruitful collaboration has been established between the ESPaCE consortium and  the MaRS radio-science and the HRSC camera teams of the MEX mission, in preparation of the December 2013 close approach of Phobos.

On the 29th of December 2013, Mars Express (MEX) came as close as 45\,km to Mars' moon Phobos -- closer than ever before. 
To guarantee the highest precision on Phobos' position, an astrometric observation campaign of Phobos was performed from MEX' on board camera SRC. Several tens of observations were performed and reduced using the UCAC4 star catalogue to provide strong constraints on Phobos' orbit in the ICRF reference frame (Pasewaldt et al., in prep.). New ephemerides of the Martian satellites were developed as part of the consortium. They are presented in this work.

In Section \ref{sec:2}, we detail the numerical model we used. In Section \ref{sec:3}, we present the whole set of observations that was used in the present work. The next section shows the astrometric results and provides physical parameters of interest like the amplitude of Phobos' forced libration and Mars' tidal $k_2/Q$ ratio. In section \ref{sec:5} we discuss an issue we encountered when considering Mars Reconnaissance Orbiter (MRO) observations. The last section is devoted to precision of the extrapolation and comparison with former ephemerides.

\section{Modeling}\label{sec:2}

We used a very similar approach as described in \citet{2007A&A...465.1075L}.
In particular, we used the NOE numerical code to model the orbits of Phobos and Deimos. The Mars rotation model considered was the one of \citet{2016Icar..274..253K} with the Mars gravity field MRO120D that we truncated at degree and order 12. \textcolor{black}{Table \ref{tab:MarsGrav} provides the numerical values of these coefficients.} The masses of \textcolor{black}{Mars ($\unit[42828.3750104]{km^3/sec^2}$)}, Phobos ($\unit[7.11 \times 10^{-4}]{km^3/sec^2}$), Deimos ($\unit[9.46 \times 10^{-5}]{km^3/sec^2})$ and the Mars Love number $k_2$ (0.169) were borrowed from this gravity solution. The planetary ephemerides were INPOP19a \citep{2020MNRAS.492..589F} that introduced the perturbations of the Sun, the Moon and all the other planets.

\begin{table*} 
\caption{\textcolor{black}{Mars gravity field used in the present work (MRO120D  (\cite{2016Icar..274..253K}) truncated at degree and order 12. Coefficients are normalised and Mars' radius is equal to 3396.0 km.}}
\begin{center} 
\begin{tabular}{lr} 
\hline 
\hline 
Harmonic coefficients&\\
\hline
$C_{2,0}, C_{3,0}, ...C_{12,0}$ &  -0.87502209245370E-03  -0.11897015037300E-04   0.51290958301340E-05  -0.17267702404260E-05  \\
& 0.13463714866160E-05   0.10598968092880E-05   0.14437523554120E-06  -0.28755906339710E-06  \\& 0.72663209592840E-06  
-0.26620541678020E-06   0.26016282936860E-06 \\
$C_{21}, C_{31}, ..., C_{12,1}$ &   0.40223333063820E-09   0.38049981991010E-05   0.42163911582170E-05   0.48384215630660E-06   \\
&0.18023583692130E-05   0.13749994266810E-05  -0.13253837279380E-06   0.42096311969260E-06  \\& 0.92399054314190E-06  
-0.81684431270170E-06  -0.11268252543490E-05 \\
$C_{22}, C_{32}, ... , C_{12,2}$&    -0.84633026559830E-04  -0.15947431923720E-04  -0.95306695299840E-06  -0.42981760456790E-05 \\  &0.86171342848250E-06   0.28139783170420E-05   0.18104244435100E-05   0.11387318459420E-05  \\& 0.69959052072370E-08 -0.31359320991050E-06  -0.97535431722040E-07 \\
$C_{33}, C_{43}, ..., C_{12,3}$ &   0.35056298360330E-04   0.64568519841300E-05   0.33126670085550E-05   0.95567075435960E-06   \\
&0.88048981361710E-06  -0.12070336086800E-05  -0.99365116894820E-06  -0.29903288633400E-06 \\ & -0.13030274194510E-05  
-0.14420200318390E-05\\
$C_{44}, C_{54}, ..C_{12,4}$ &  0.30824936247700E-06  -0.46407608474120E-05   0.10087553624920E-05   0.24689899049820E-05  \\
& 0.15882837683240E-05   0.29497879998410E-06  -0.12139330875140E-05  -0.15741226303010E-05 \\ 
& -0.10141072962050E-06\\
$C_{55}, ...C_{12,5}$ &  -0.44492645268970E-05   0.16578866158220E-05  -0.19168169008940E-06  -0.27917524692450E-05 \\
& -0.22719046508670E-05   0.42141118528680E-06   0.13564965343170E-05   0.72031532166240E-06 \\
$C_{66}, ... C_{12,6} $&   0.27622296666680E-05  -0.55960104736480E-06  -0.91440928257930E-06   0.80860626285270E-06 \\
&  0.66960055718320E-06  -0.24475256306890E-06  -0.40343909794620E-06\\
$C_{77}, ... C_{12,7} $&   0.44039558349960E-06  -0.47364667411900E-06  -0.61795403730490E-06   0.31785454600440E-06  \\
& 0.66045729925750E-06   0.40444336054540E-06\\
$C_{88}... C_{12,8} $&  -0.31065725923520E-06   0.12080798860860E-05   0.54059795600680E-06  -0.11820682984680E-05 \\
& -0.16039741594790E-05\\
$C_{9,9}, ...C_{12,9}$ &  -0.11719047495280E-05  -0.14575391837510E-05  -0.41520987784620E-06   0.70845500937310E-06\\
$C_{10,10}...C_{12,10} $& -0.27477718238890E-06   0.33329692241000E-06   0.48850428613000E-06\\
$C_{11,11}, C_{12,11}$ &   -0.44950881784560E-07   0.86475982548170E-06\\
$C_{12,12}$ &   -0.66853201533900E-08\\

$S_{21}, S_{31}, ..., S_{12,1}$ &    0.23031838535520E-10   0.25177117707630E-04   0.37632643561220E-05   0.21231129753950E-05   \\
&  -0.15185193991260E-05  -0.22737194866050E-06   0.75052081765850E-06  -0.49021207335710E-06  \\
&  0.22320069479760E-06  -0.22130696733310E-06  -0.51245929517270E-07 \\
$S_{22}, S_{32}, ... , S_{12,2}$&   0.48939418321670E-04   0.83623939784670E-05  -0.89807968418080E-05  -0.11656954440860E-05    \\
  &  0.14691007371520E-05  -0.62967694381360E-06   0.50705823024380E-06   0.38219982478050E-06  \\
& -0.11157916028810E-05  -0.99401812075920E-06   0.53304503764570E-06\\
$S_{33}, S_{43}, ..., S_{12,3}$ &   0.25571325457370E-04  -0.19377212284160E-06   0.27144097785790E-06   0.33292558689320E-06    \\
& -0.39698286606070E-06  -0.13413009168640E-05  -0.10076169517160E-05   0.44355180981810E-06 \\ 
&  0.76528953740340E-06   0.31477016571890E-06 \\
$S_{44}, S_{54}, ..S_{12,4}$ &  -0.12873056977380E-04  -0.33815536222490E-05   0.26386569471530E-05  -0.42245529297430E-06    \\
&  0.14812074290120E-06   0.16235822767760E-05  -0.69716893597150E-07  -0.64448534121760E-06  \\ 
&  0.12358719897030E-06\\
$S_{55}, ...S_{12,5}$ & 0.37804789409520E-05   0.16226764849180E-05  -0.13585219042100E-05  -0.16297926112810E-05   \\
&   -0.15493068736430E-05  -0.10629285253530E-05   0.89908740763190E-06   0.10016903702080E-05\\
$S_{66}, ... S_{12,6} $&   0.82135333243850E-06  -0.19013643905730E-05  -0.17899324707530E-05   0.57773114546310E-06  \\
&  0.11171248335330E-05   0.20644450708130E-07  -0.16298585248040E-05 \\
$S_{77}, ... S_{12,7} $&   -0.17756701426830E-05   0.16446964596680E-05   0.86891992653820E-06  -0.62023200497360E-06    \\
& -0.86497561721440E-06  -0.11283348722240E-06 \\
$S_{88}... S_{12,8} $&  -0.25028184874420E-06  -0.14644731038890E-06   0.82057425718930E-06   0.76703348424800E-06  \\
&  -0.38794394344050E-06\\
$S_{9,9}, ...S_{12,9}$ & -0.65779217295200E-06  -0.14635579222260E-05  -0.41944370089590E-06   0.47674076749600E-06  \\
$S_{10,10}...S_{12,10} $& 0.75328042634720E-06   0.19666657150790E-05   0.13782862550090E-05 \\
$S_{11,11}, S_{12,11}$ & -0.32345607088980E-06  -0.16550385123550E-05  \\
$S_{12,12}$ & -0.89210620377340E-07 \\

\hline
\end{tabular} 
\end{center} 
\label{tab:MarsGrav}
\end{table*}

Three amendments on the dynamics were introduced in comparison to \citet{2007A&A...465.1075L}. The first two were the introduction of general relativity and tidal-cross effects \citep{2017Icar..281..286L} that consist in the cross action of the tidal bulges raised on Mars by different tidal raising bodies. While rather negligible at the current level of accuracy for Mars, we added these effects for completeness. Tides on Mars' moons were neglected.

A third but important dynamical effect was the introduction of the forced libration on the rotation of Phobos (\citet{2010AJ....139..668J}). Indeed, such perturbation introduces a secular effect on the periapsis of Phobos (\citet{1990AA...233..235B}, \citet{2019Icar..326...48L}), barely maskable in the fitting procedure. The secular drift, associated with the quadrupole field of Phobos on its periapsis (or any moon in somewhat similar configuration) was given by \citet{1990AA...233..235B}
and recalled by \citet{2010AJ....139..668J}
to be
\begin{eqnarray}
\Delta \varpi&=&\frac{3}{2}\left(\frac{R}{a}\right)^2\left[J_2-2c_{22}\left(5-\frac{4A}{e}\right)\right]nt\nonumber\\
&&+\frac{3}{2}\left(\frac{R}{a}\right)^2\frac{(J_2+6c_{22})}{e}\sin(\cal{M})\label{eq:libr}
\end{eqnarray}
where $a, e, n, \cal{M}$ and $\varpi$ are the traditional osculating Keplerian semi-major axis, eccentricity, mean motion and mean anomaly while $R, J_2=-c_{20}$ and $c_{22}$ denote the mean radius and un-normalized gravity coefficients of Phobos. Here we explicitly see the action of the libration amplitude $A$ on the periapsis. Our $A$ is $-A$ in \citet{2010AJ....139..668J}. Of course, this action of the quadrupole field of the satellite is just one effect affecting the pericenter. The precession of the pericenter is mainly due to the flattening of Mars.

Since the gravity field of Phobos is hard to determine directly from the current data \citep{2019MNRAS.490.2007Y},
we relied on \citet{2014PSS..102...51W} 
who used Phobos' shape and homogeneous density hypothesis to derive Phobos' gravity field. 
They obtained for the first gravity field coefficients $c_{20}=-0.066127$ and $c_{22}=0.009917$, assuming a Phobos' radius of 14.0$km$.

\section{Observation sets}\label{sec:3}

Most data sets used in this work were already processed in \citet{2007A&A...465.1075L} and \citet{2010AJ....139..668J}, and benefited greatly from the astrometric catalogue of \cite{1989AAS...77..209M}.  A large astrometric database is available on the Natural Satellites DataBase (NSDB) server \citep{2009AA...503..631A}. It turned out that some of the old measurements were extremely biased and therefore were eventually not included in the fit. Moreover, ground observations with residuals larger than 2 arcseconds were removed. A rejection criteria beyond $3\sigma$ was applied to ground observations.
Last, spacecraft data were weighted in the same way as in \citet{2019Icar..326...48L}. Here we present two important new data sets, not available to the former works.

\subsection{Observations from digitized plates}

Following the discovery of Phobos and Deimos, and the detection of the long suspected secular accelerations in their longitudes  \citep{1945AJ.....51..185S}, the United States Naval Observatory (USNO) began a thirty-year (1967-1997) program of photographic observations of the Martian satellites \citep{1977plsa.conf...63P,1978VA.....22..141P,1979nasm.conf...17P,2012pascunaroo}. These observations were among the most accurate \citep{1989AAS...77..209M} and were used to support all space reconnaissance projects of the Martian system.

Photographic observations were begun in 1967 and continued at every opposition through 1997. They were taken with the USNO 61-inch astrometric reflector in Flagstaff, Arizona, and the USNO 26-inch refractor in Washington, D.C. with the use of special filters. Several Schott 5 inch x 7 inch x 3 mm GG14 (yellow) filters were polished optically flat. In the center of each a small, thin metallic nichrome film having an optical density of about 3.0 was deposited by evaporation. The GG14 was chosen to accommodate both telescopes, and nichrome was chosen because it transmits neutrally in the visual bandwidth. The function of the small nichrome filter was to reduce the intensity of the planetary image to that of the satellites, producing a measurable image of the planetary disk. A number of Kodak emulsions were used, including 103aJ, 103aG, and IIIaJ. More details about the observations and astrometric results are available in \citet{2015AA...582A..36R}.

Four hundred twenty-five plates were selected and transmitted to Royal Observatory of Belgium (ROB) to be digitized \citep{2011MNRAS.415..701R,2011ASPC..442..301D}. Each plate contains two to three exposures shifted in the RA direction. Measured $(x,y)$ plate positions were corrected for instrumental and spherical effects, and the reductions were performed using four or six suitable constant functional models \citep{2011MNRAS.415..701R,2014A&A...572A.104R} to provide equatorial (RA, DEC) astrometric positions of the planet and its satellites. The digitizations and measurements resulted in 777 positions of Mars, 640 positions of Phobos, and 704 positions of Deimos.

The observed positions of Mars, Phobos, and Deimos were compared with their theoretical computed positions given by DE430 planetary ephemeris \citep{2014IPNPR.196C...1F} and NOE MarsSatV1\_0 satellite ephemerides \citep{2007A&A...465.1075L}. The key point was that the NOE MarsSatV1\_0/DE430 astrometric residuals (O-C) for all observations had an rms of 47.8 mas, 60.5 mas and 50.7 mas, for Mars, Phobos and Deimos, respectively. \textcolor{black}{Overall intersatellite RMS (O-C) is 39 mas.} These rms correspond to the observation accuracies over thirty years, providing observations that are similar in accuracy to old space data but with a more Gaussian profile and a larger time span.

\subsection{New MEX/SRC data}\label{subsection:MEX/SRC}

MEX is in a highly elliptical and nearly polar orbit about Mars reaching well beyond the almost circular path of Phobos \citep{jau07}. Due to the asphericity of the planet's gravitational field its longitude of ascending node is drifting westwards and its argument of pericenter is drifting in the direction opposite to the spacecraft motion. At the same time its orbital period is similar to the one of the inner satellite of Mars resulting in several flyby opportunities in consecutive orbits at intervals of about five to six months. Whenever a close encounter between the space probe and this Martian moon occurs high resolution observations of Phobos by the onboard cameras are planned, see section \ref{sec:close encounters}. In the past years new opportunities to observe Phobos and Deimos in conjunction with other solar system objects or stars (so-called mutual events, see section \ref{sec:mutual events}) were identified and more regularly planned. These observations do not necessarily require the proximity to the body \citep{2018A&A...614A..15Z}.

The Super Resolution Channel (SRC) is part of the High Resolution Stereo Camera (HRSC) onboard MEX. It features a compact Maksutov-Cassegrain optical system, a 1K by 1K interline-transfer CCD, and fast read-out electronics \citep{obe08}.

\subsubsection{Close encounters} \label{sec:close encounters}

During an approach maneuver the camera is pointed to a fixed location on the celestial sphere and is rotated around the boresight axis such that the lower and upper image borders are parallel to the relative velocity vector of Phobos w.r.t. the spacecraft.

In the usual eight-image sequences the first and the last pictures are long-time exposures to detect the faint light of background stars. Exposure times of the images in between are adjusted to the brighter Phobos surface. Pointing corrections in sample and line are derived from star images and then interpolated linearly for the pictures of Phobos \citep{wil08}.

For the purpose of astrometry, SRC images have been focussed using an image-derived point spread function in a Richardson-Lucy deconvolution \citep{mic09}. The image positions of stars have been determined by fitting a 2D Gaussian profile to the pixel value distributions (Duxbury 2012, pers. comm.). Pointing corrections have been derived w.r.t. the predictions of reference star positions based on the UCAC4 catalog. The directions to Phobos have been measured by applying the limb-fit approach based on the latest 3D shape model by \citet{wil14}.

Between May 2013 and March 2014 a larger number of flybys than usual have been performed in order to support the Phobos gravity field experiment. Therefore data from 38 approach maneuvers with distances ranging from 350 to 14\,000 km have been evaluated. Summing all up, 340 astrometric observations of Phobos in right ascension and declination coordinates have been provided. Estimated uncertainties vary from hundreds of meters up to one km (Pasewaldt et al., in prep.).

Hitherto, the estimation of uncertainties is still incomplete. This is due to the occasional presence of spacecraft jitter during image acquisition (see \citet{2015AA...580A..28P} for details). For sequences with continuous observations of reference stars or other celestial objects this causes no problem. But for several eight-image series with only two star pictures an additional uncertainty is introduced that is only hard to quantify. Therefore, we distinguished our Phobos measurements into those performed during linear variation and those carried out during non-linear variation in camera pointing.

Besides the measurement of reference objects, the CCD line coordinate of a Phobos position could provide information on the presence or absence of spacecraft oscillations during image recording. Based on the imaging geometry described above this should be almost constant throughout a sequence. From at least three successive Phobos observations we can calculate two line coordinate differences.

While only positive or negative differences have been classified as linear variations, alternating positive and negative differences have been categorized as non-linear variations in pointing. If in addition the pre-fit residuals were deviating by more than three sigma from the arithmetic mean, we marked this as an outlier. For series with only two Phobos measurements we could not determine whether the outlier has been related to a non-linear change in pointing or not. Further on, in some images we fitted the shape model-derived limb to only very short limb point arcs. 

This classification is not an equivalent substitute for a proper weight estimate, but it gives at least a suggestion for the following evaluation of our data.

\subsubsection{Mutual events observations} \label{sec:mutual events}

Longer imaging sequences of mutual events showing either both Martian moons or one of the moons with Jupiter or Saturn in the background were analyzed by \citet{2018A&A...614A..15Z}. 
These observations can be obtained more frequently than direct observations during flybys.
Moreover, a wide range of MEX orbit positions are possible. However, in contrast to close encounter images, Phobos and Deimos  can also cross the image plane diagonally. It also occurs that the moons are only partially visible.

To determine the bodies' locations within the image, simulated images are computed.
The simulations are based on the ephemerides model mar097, a rotational model by Stark et al. (see \cite{2018CeMDA.130...22A} and \cite{2019CeMDA.131...61A}) and shape models of Phobos and Deimos derived by \citet{wil14} and \citet{Thomas-etal_2000}, respectively. To achieve a better agreement with the observation, the simulation is convolved with a point spread function describing the image distortion of the SRC. Here a subset of the PSF derived by \cite{Duxbury-2011} was applied.
Matching the illuminated simulation against the observation provides the bodies’ line and sample coordinates with sub-pixel accuracy. For images showing either Phobos and/or Deimos with a far distant object,
the derived right ascension and declination of the moon (as seen from MEX) are determined.
The exhaustive list of MEX data we used is given in Table \ref{tab:moysig}. \textcolor{black}{In particular, due a current limitation of our software, mutual events evolving both moons simultaneously (i.e. without the presence of Jupiter or Saturn) have not been included in the current solution.}

\section{Adjustment results}\label{sec:4}

\textcolor{black}{All data were fitted using the RMS of each data set as a priori uncertainty. Data points higher than 2 arcsec for ground observations were systematically removed. Then, a 3 sigma rejection criteria was applied. For spacecraft observations, we used the published estimated precision. However, to assess a better weight, we rescaled these uncertainties by a scalar chosen to get about 66\% of astrometric residuals within 1$\sigma$ level for each satellite and coordinate.}

\begin{figure*} 
\begin{center} 
\begin{tabular}{ll} 
\hspace*{-0.35cm}\includegraphics[width=9.5cm,angle=0]{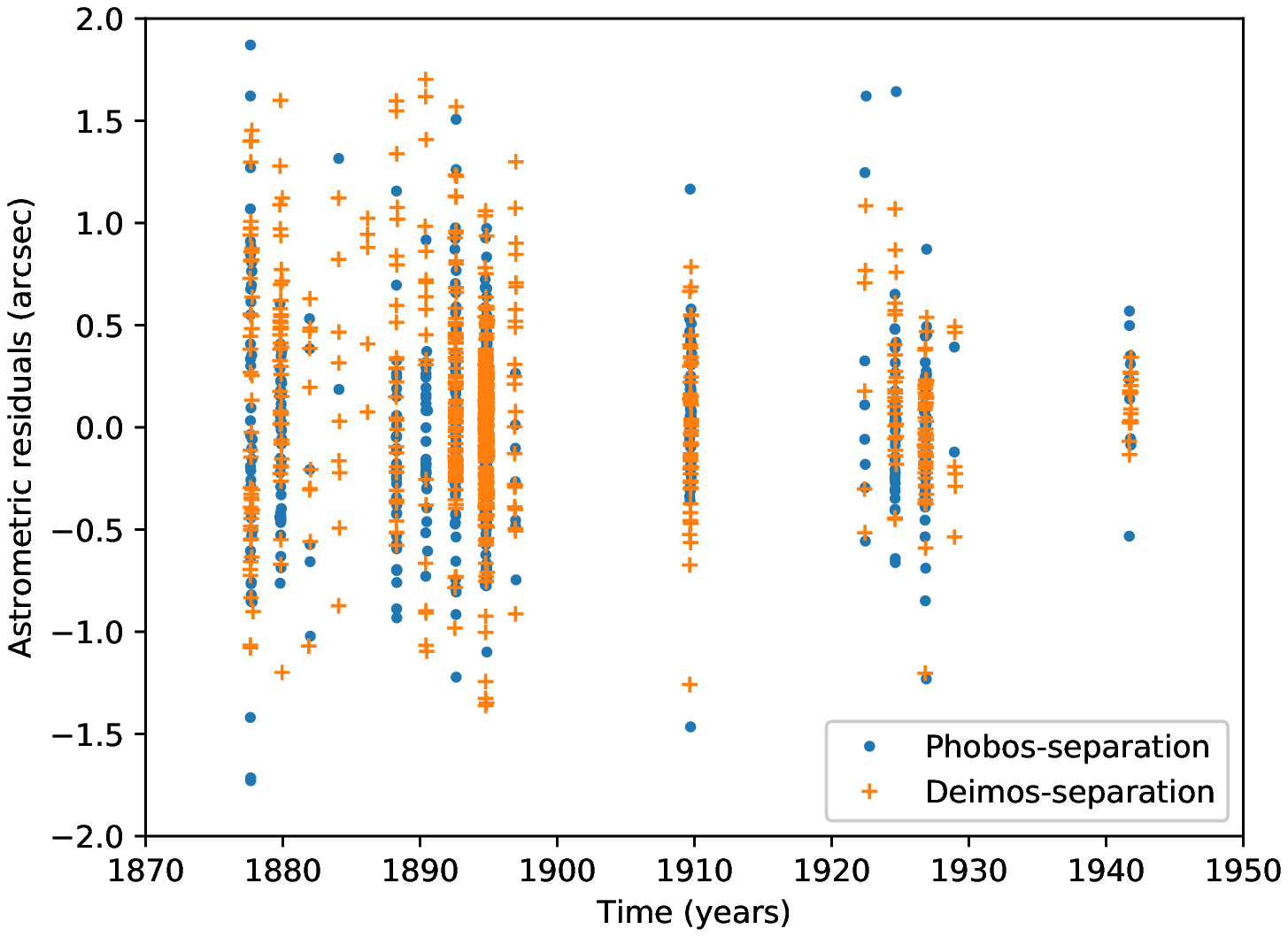} & \includegraphics[width=9.5cm,angle=0]{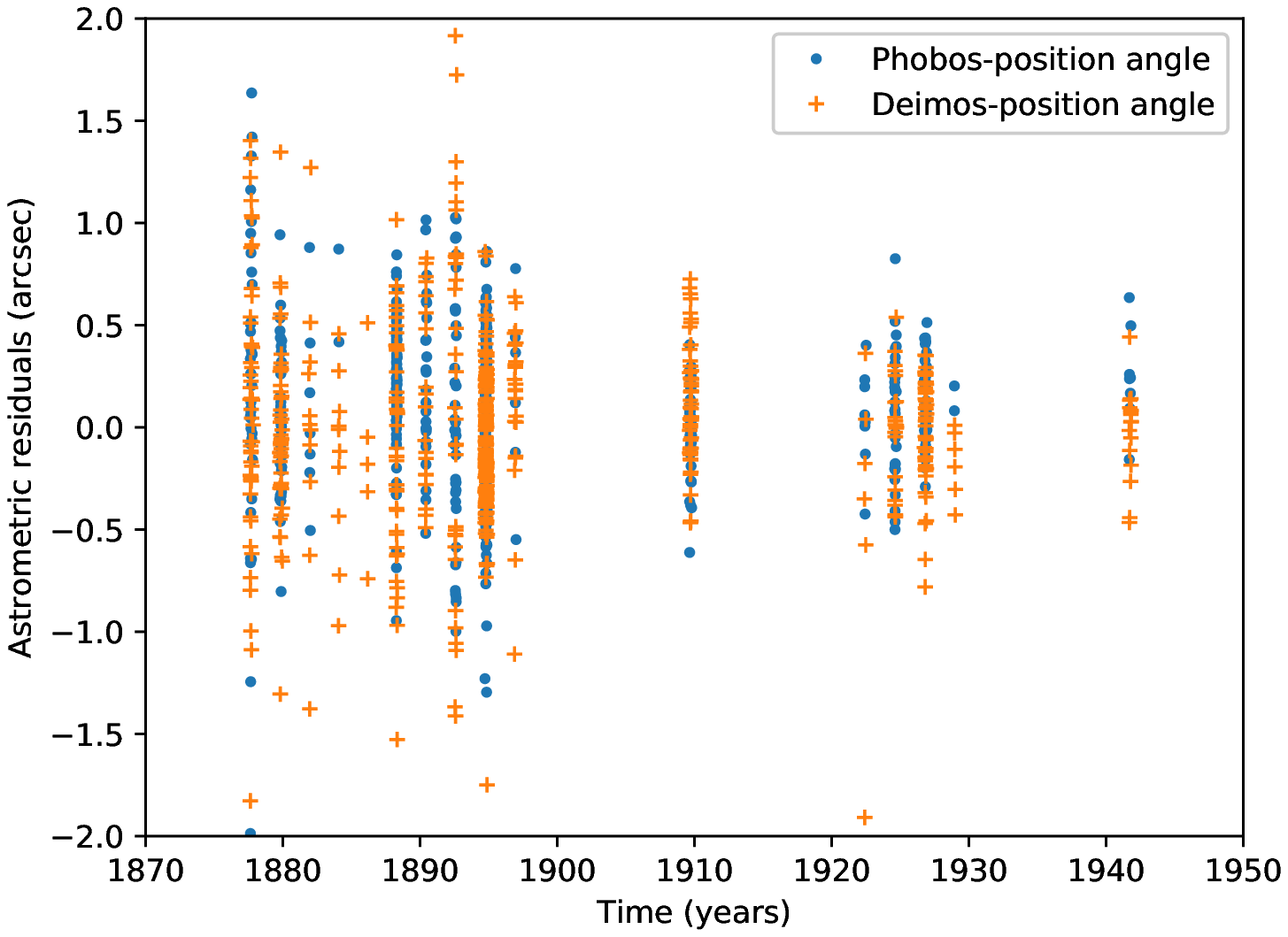}\\
\hspace*{-0.35cm}\includegraphics[width=9.5cm,angle=0]{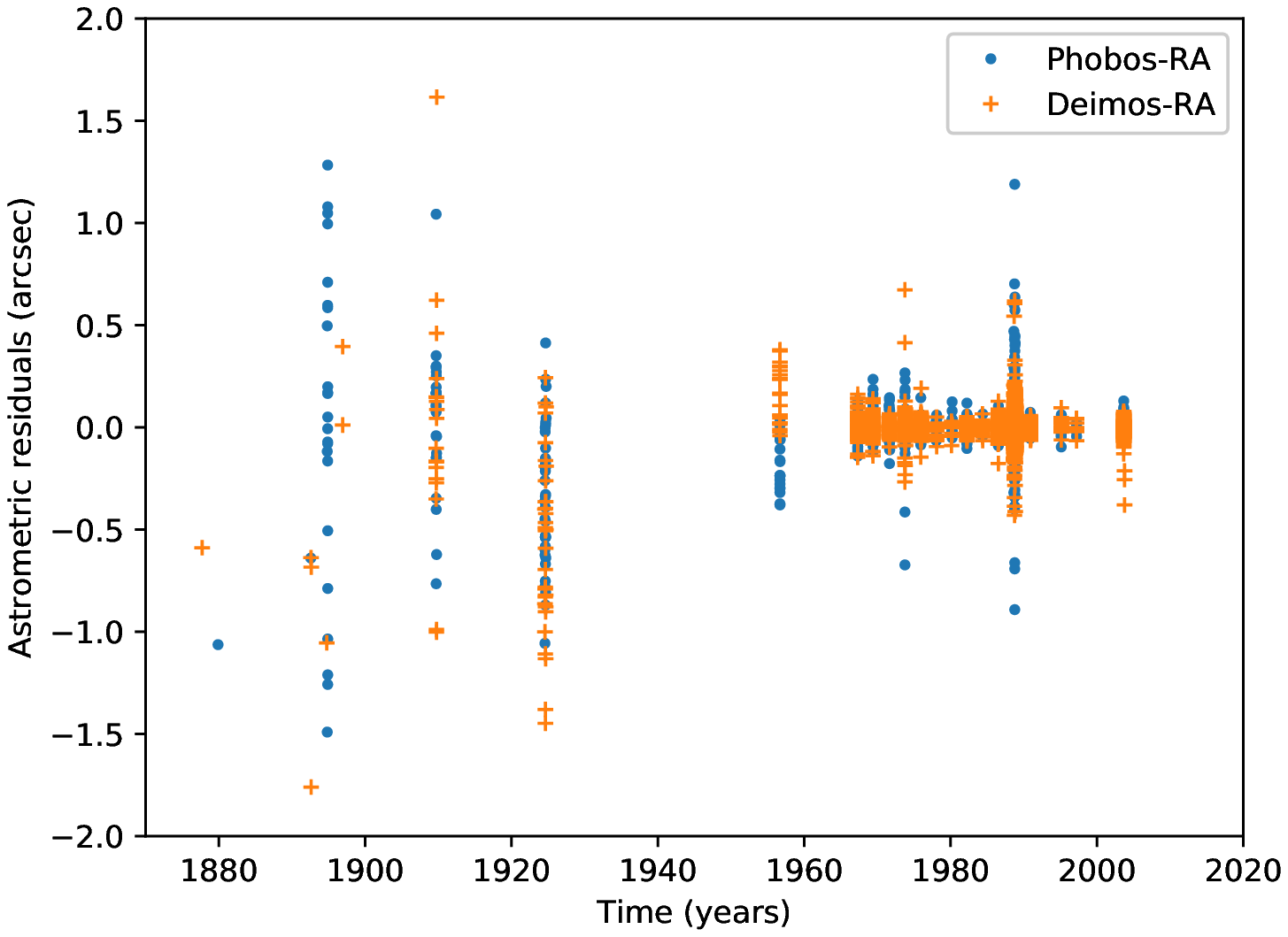} & \includegraphics[width=9.5cm,angle=0]{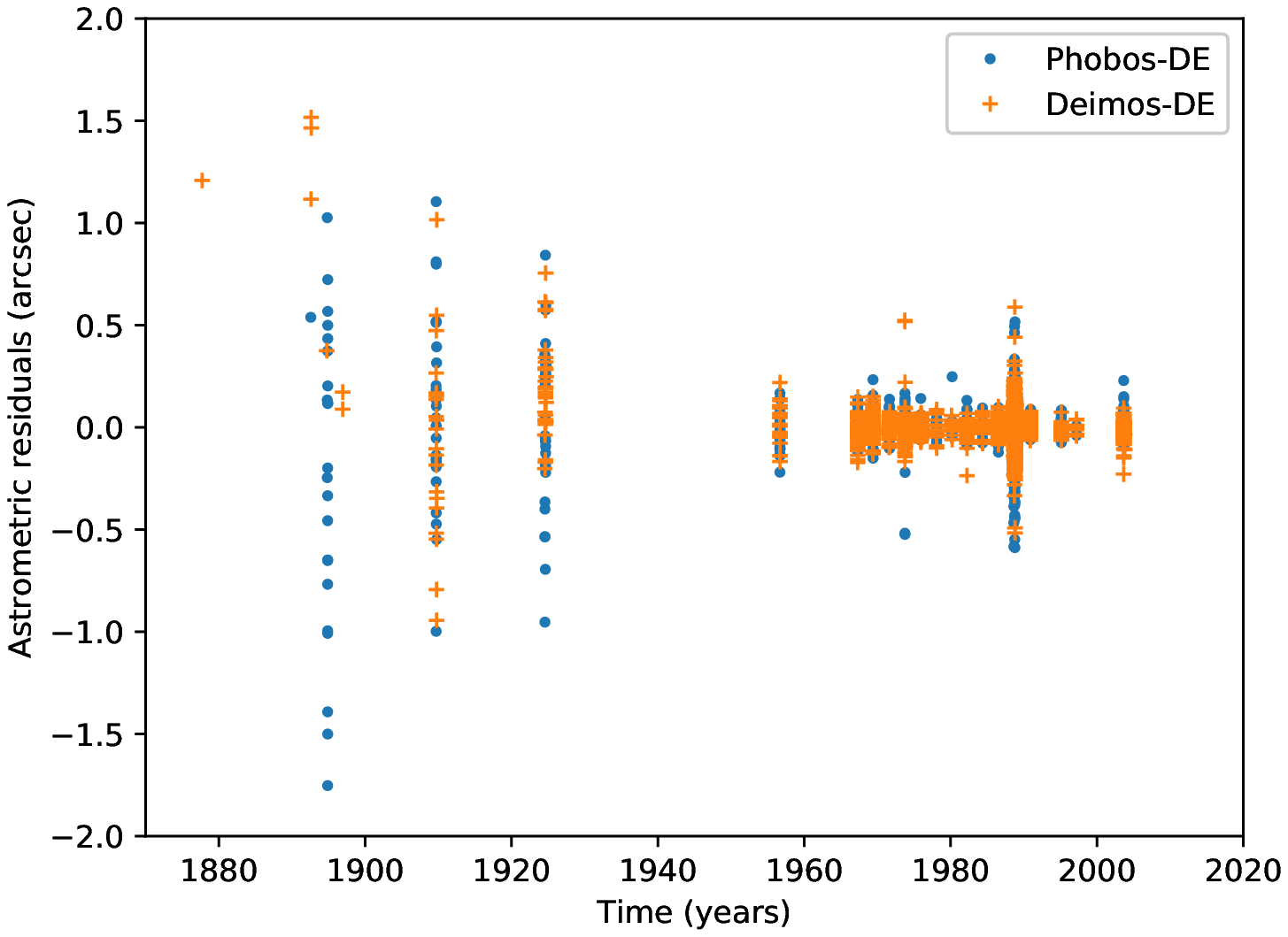}\\
\end{tabular} 
\caption{Astrometric residuals after fit between the model and the ground observations for Phobos and 
Deimos (top: separation and position angle data, bottom: right ascension and declination data). The satellites' initial positions and velocities, the Martian dissipation quality factor $Q$ and Phobos' forced libration amplitude were fitted here. The position angle was multiplied by the separation to provide residuals in arcsec.}\label{fig:earth-res}
\end{center} 
\end{figure*}

\begin{figure*} 
\begin{center} 
\begin{tabular}{ll} 
\hspace*{-0.35cm}\includegraphics[width=9.5cm,angle=0]{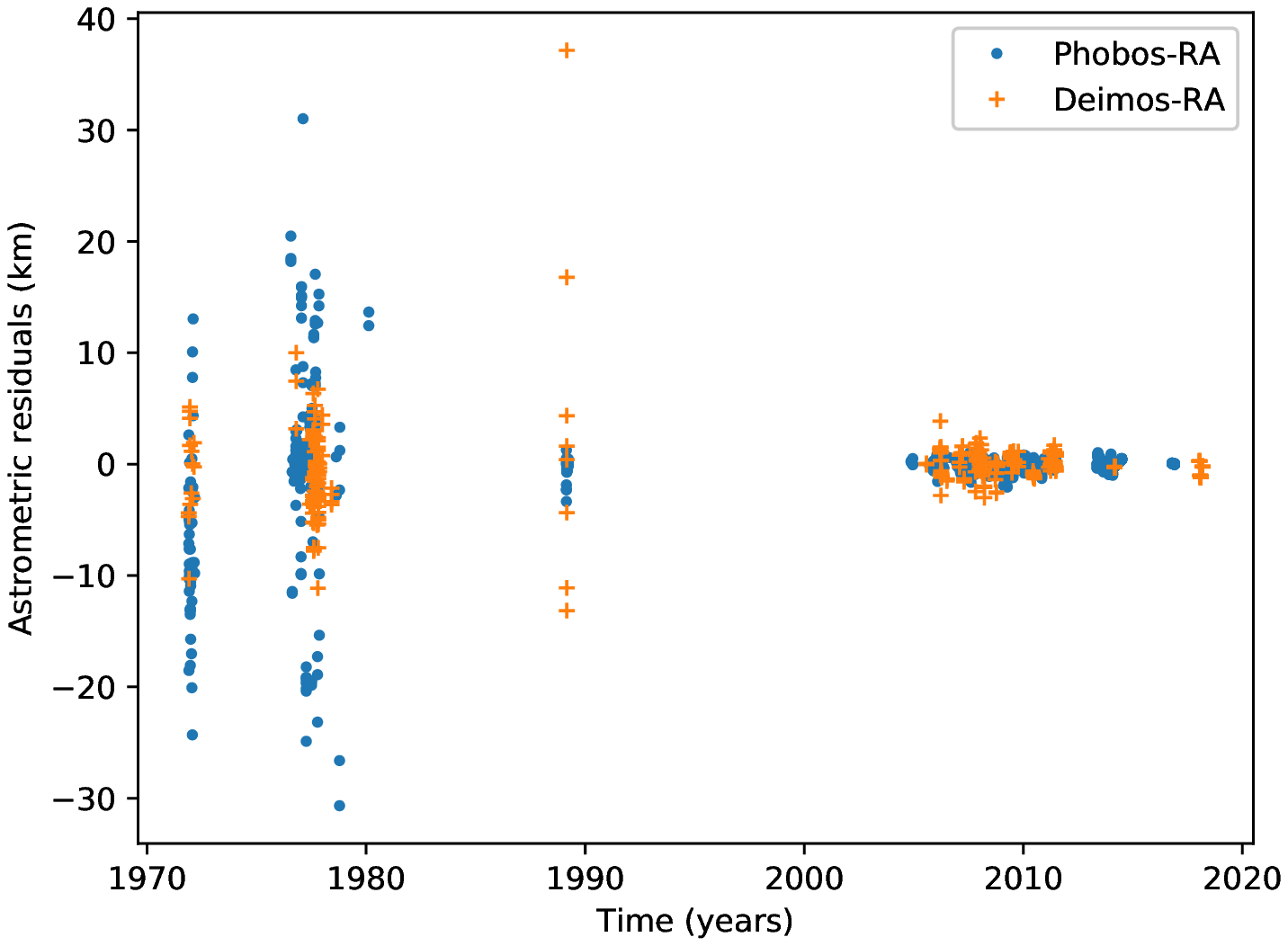} & \includegraphics[width=9.5cm,angle=0]{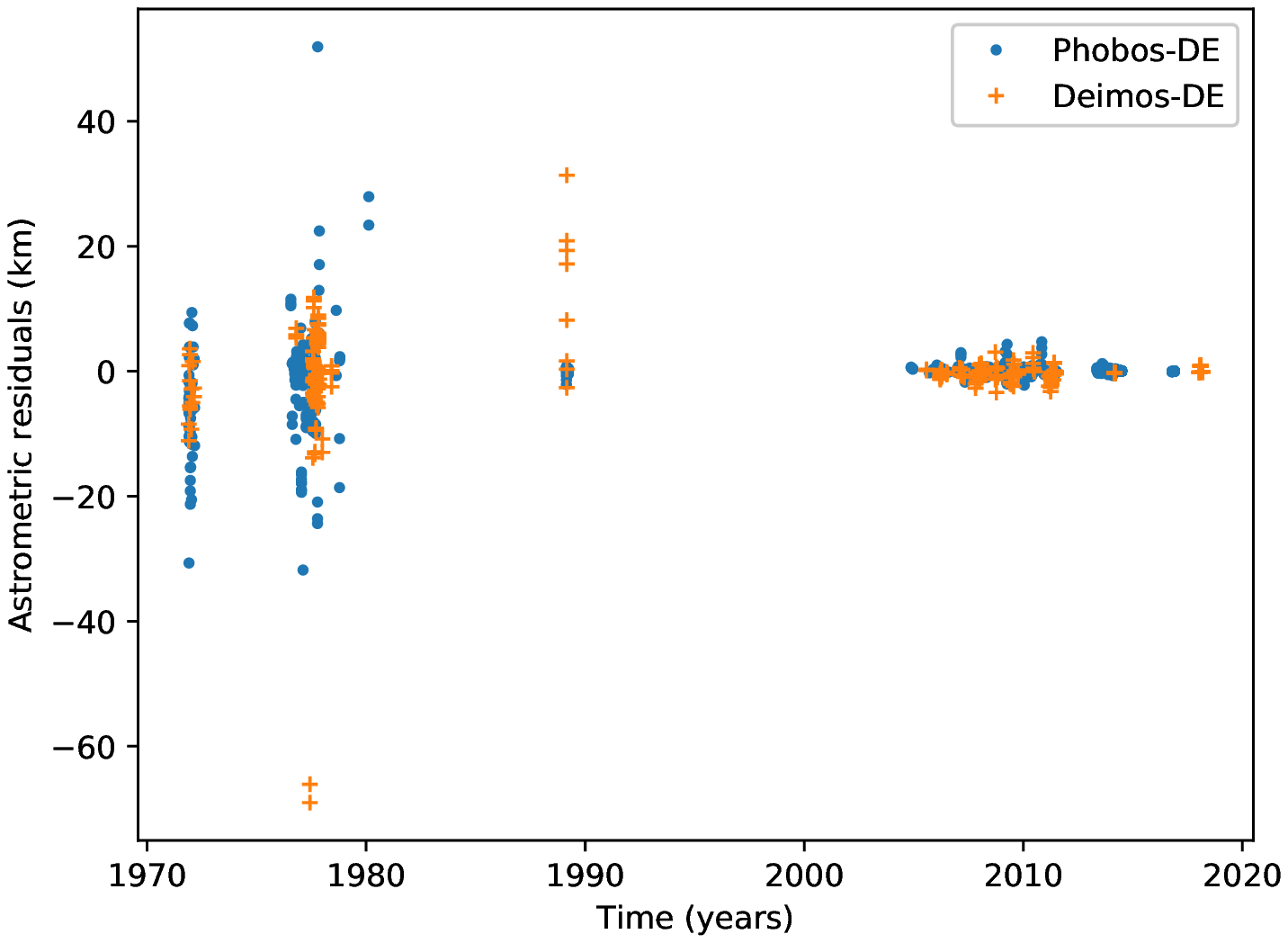}
\end{tabular} 
\caption{Differences in distance after fit between the model and the spacecraft observations for Phobos and Deimos (left: right ascension; right: declination). The satellites' initial positions and velocities \textcolor{black}{(at epoch J2000 i.e. Julian day 2451545.0)}, the Martian dissipation quality factor $Q$ and Phobos' forced libration amplitude were fitted here. 
\label{fig:SC-observations}}
\end{center} 
\end{figure*}

\begin{figure*} 
\begin{center} 
\begin{tabular}{ll} 
\hspace*{-0.35cm}\includegraphics[width=9.5cm,angle=0]{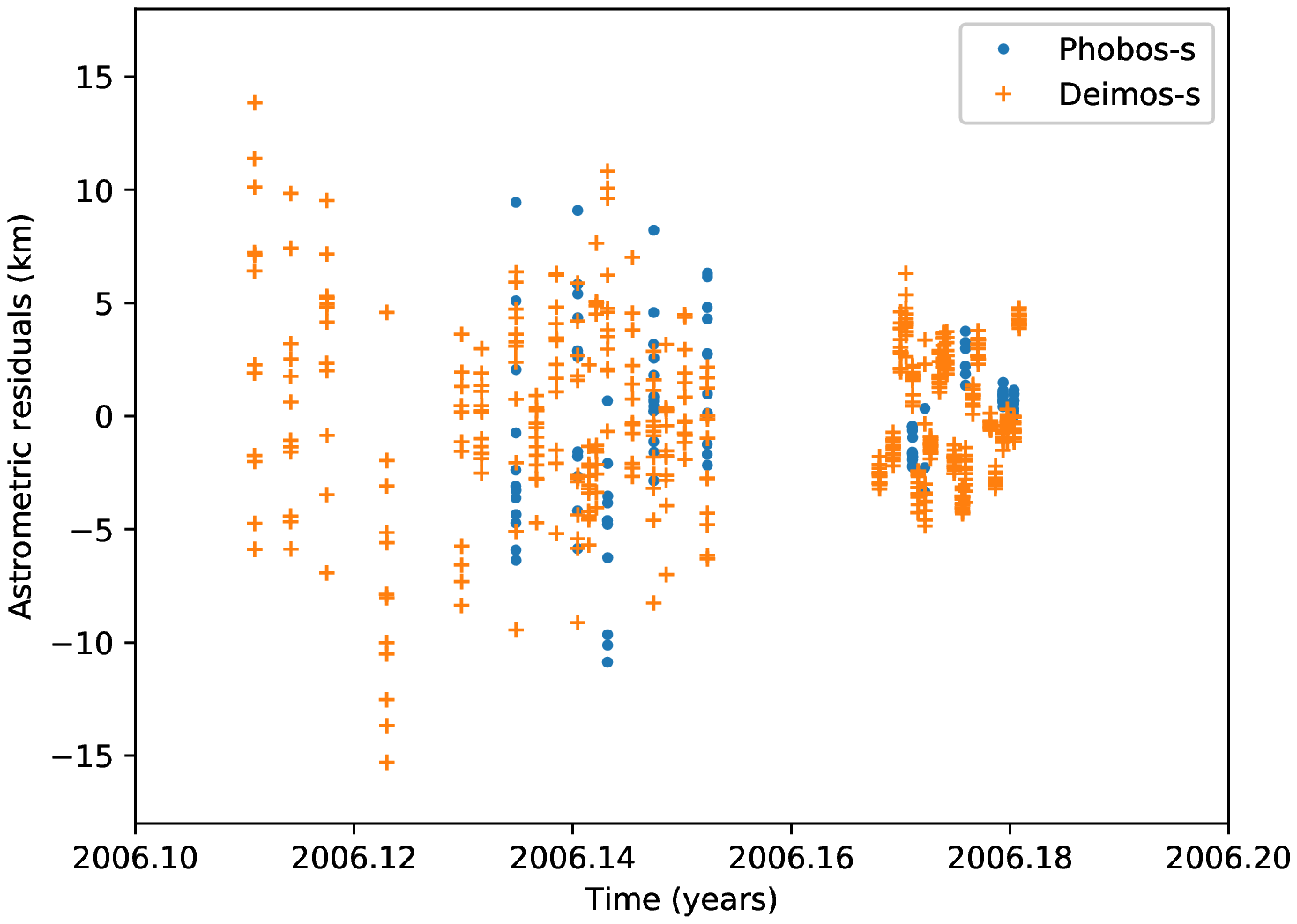} & \includegraphics[width=9.5cm,angle=0]{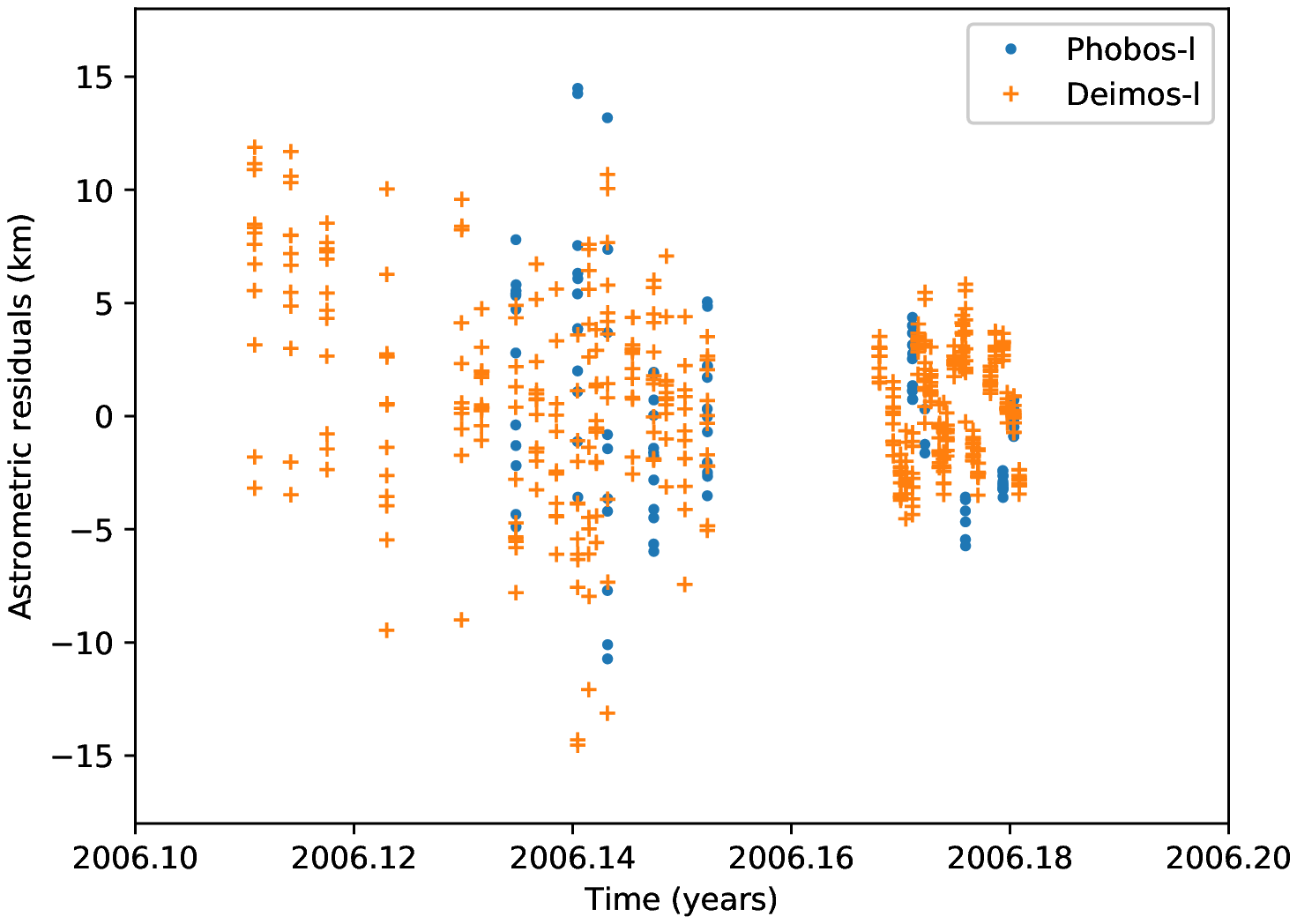}\\
\hspace*{-0.35cm}\includegraphics[width=9.5cm,angle=0]{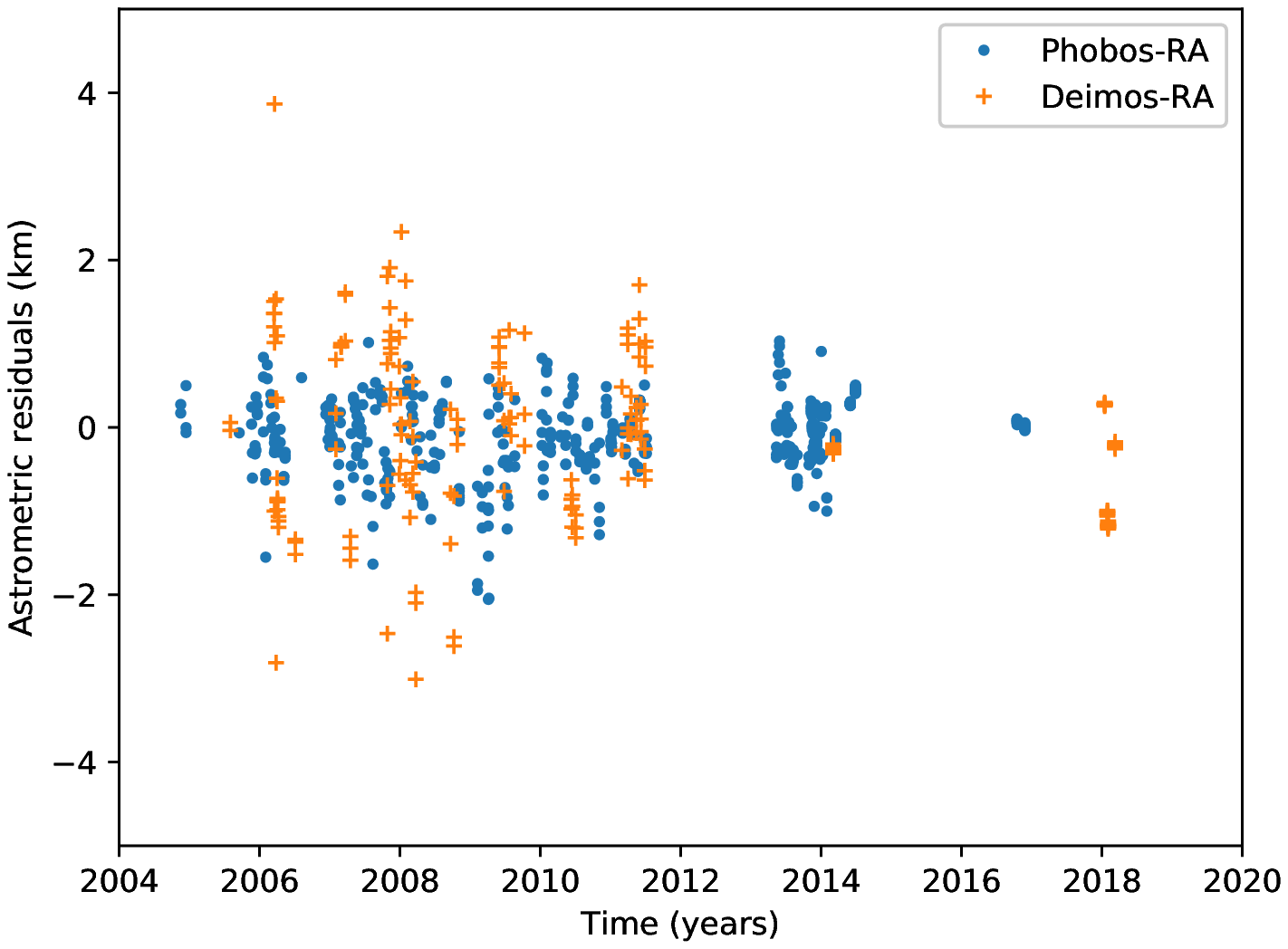} & \includegraphics[width=9.5cm,angle=0]{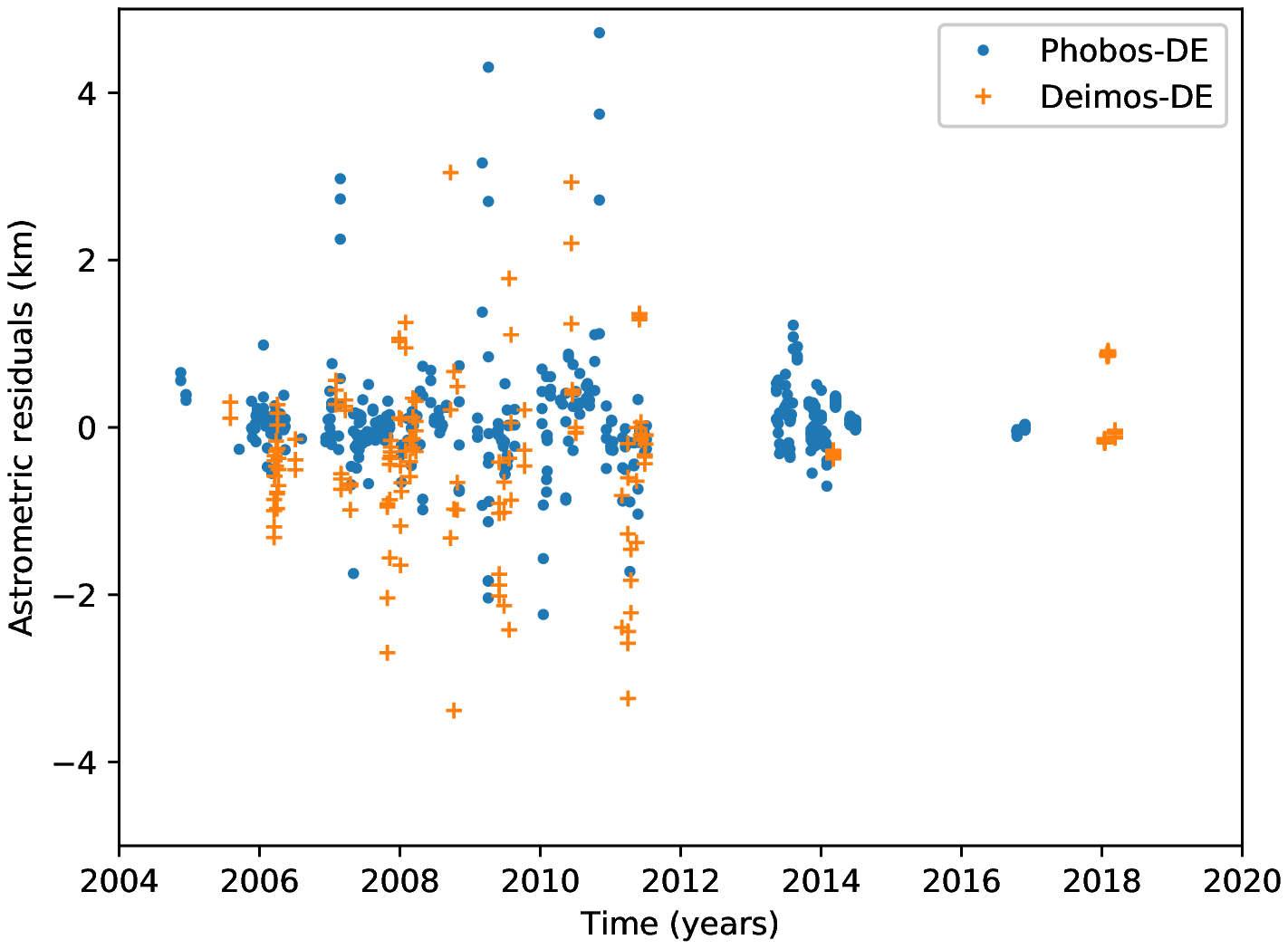}
\end{tabular} 
\caption{Differences in distance after fit between the model and: i) the MRO data (top); ii) the MEX data (bottom). The satellites' initial positions and velocities, the Martian dissipation quality factor $Q$ and Phobos' forced libration amplitude were fitted here. }\label{fig:MRO-MEX}
\end{center} 
\end{figure*}

We provide in Figures \ref{fig:earth-res}-\ref{fig:MRO-MEX} and Tables \ref{tab:moysigSP}-\ref{tab:CI} the astrometric residuals of all ground and space data used in the development of the present ephemerides \textcolor{black}{as well as initial conditions of Phobos and Deimos after fit}. The high precision of both the reduction of photographic plates and the new MEX data is confirmed, with typical residuals at the level of 40-50 mas and 300 meters, respectively. 
\textcolor{black}{The combination of the existence of two Mars moons with the use of inter-satellite fitting often provides both opposite mean values and similar sigma. Such feature breaks up when both moons could not be observed always simultaneously, as indicated by a different number of observation per moon.}
We note that the MEX residuals may be associated with an error on the astrometric calibration of the field and the error on the spacecraft position. In particular, it was found that the long imaging sequence of mutual events (see subsection \ref{subsection:MEX/SRC}) introduced a significant bias in the fitting procedure. While the jitter effect of the spacecraft could be removed thanks to the mutual event opportunity, the error on MEX' position in space remained. An error of the order of a few hundred of meters on MEX' reconstructed orbit was found, much higher than the initially estimated error at the level of 20-25 meters \citep{2008P&SS...56.1043R}. It appeared that the increase of the error on MEX' position in 3-dimensional space arose from an unfortunate consequence of the new tracking strategy of MEX. Indeed, since 2011 ESA has decreased the amount of tracking data around periapsis, entailing a poorer constraint on MEX' motion.

Thanks to the new sets, we formally gain an order of magnitude on the uncertainty (1$\sigma$) of the Martian tidal quality factor and the physical libration of Phobos. 
Assuming $k_2$=0.169, we obtained $Q=93.68 \pm 0.18$. Since the eccentricity of Phobos' orbit is small (less than 2\%), the secular variation of its semi-major axis can be easily approximated using \citep{1964RvGSP...2..661K}
\begin{eqnarray}
\frac{da}{dt}\simeq-\frac{3k_2mnR^5}{QMa^4}, \label{eq:da/dt}
\end{eqnarray}
where $m$ and $M$ denote Phobos' and Mars' mass, respectively. From \eqref{eq:da/dt} we get for the secular acceleration on longitude 
\begin{eqnarray}
\frac{1}{2}\frac{dn}{dt}\simeq+\frac{9n^2k_2mR^5}{4QMa^5}.
\end{eqnarray}
\textcolor{black}{Using the approximation $n^2a^3\simeq GM$, and the value 9378km for Phobos semi-major axis, we obtain for Phobos' tidal acceleration $1.24 \times \unit[10^{-3}]{ deg/yr^2}$. A more accurate estimation can be determined by running the full numerical model with/without tidal dissipation inside Mars. This provides for Phobos' tidal acceleration $(1.257 \pm 0.003 )\times \unit[10^{-3}]{ deg/yr^2}$.}
In practice, this orbital decay and associated longitude acceleration is what allows us to estimate Mars' $Q$. 
It is noteworthy to recall that we do not include higher Love numbers ($k_3, k_4$...) in our fit. Hence, our estimation of $Q$ related to $k_2$ should in reality be more considered as a lumped $Q$ parameter. Considering the high precision we now get on this tidal parameter (less than 0.2\%), it would probably make sense to add at least $k_3$. Nevertheless, it would be practically impossible to solve for two different $Q$, one associated with each tidal frequency.

Phobos' physical libration was simultaneously found to be $A=1.09 \pm 0.01$ degrees (Table \ref{tab:libration}). This last value is at $5\sigma$ below its theoretical value of 1.14 degrees \citep{2014PSS..102...51W} assuming homogeneity. Nevertheless, considering the very small error bar (1\%), it is pretty clear that in the harmonics expansion of Phobos a higher order degree should be considered, as well as an improved rotation model \textcolor{black}{like suggested by \cite{2012A&A...548A..14R}}, before drawing any conclusion on Phobos' interior. 

\begin{table*} 
\caption{Mean ($\nu$) and standard deviation ($\sigma$) on separation $s$ and position angle $p$ (multiplied by the separation) in seconds of 
degrees for each satellite. $N$ is the number of observations by satellite (one number per coordinate). The year appearing next to each observatory name corresponds to the observed Mars opposition.}
\begin{center} 
\begin{tabular}{lrrrrrl} 
\hline 
\hline 
Observations&$\nu_s\ \ \ $&$\sigma_s\ \ \ $&$\nu_p\ \ \ $&$\sigma_p\ \ \ $&$N\ \ $&satellite\\
& ('')$\ \ \  $& ('')$\ \ \  $& ('')$\ \ \  $& ('')$\ \ \  $&  & \\
\hline 
\citet{1989AAS...77..209M}, Hall-Newcomb-Harkness                                    
&  -0.0130 &   0.6717 &   0.0964 &   0.5627 &    112,    110  & Phobos\\
 U.S. Naval Obs., Washington (before 1893)&   0.2381 &   0.6856 &   0.0141 &   0.5803 &    147,    155 & Deimos\\
\hline
\citet{1888AJ......8...73K}, \citet{1890AJ.....10...89K}
 &  -0.1450 &   0.3827 &   0.2262 &   0.2840 &     59,    103 & Phobos\\
Lick Observatory 1888, 1890&   0.1310 &   0.6332 &   0.0200 &   0.5633 &     33,     46& Deimos\\
 \hline
\citet{1892AJ.....12..137C}, \citet{1895AJ.....15....1C}
  &   0.0363 &   0.2243 &   0.0375 &   0.4388 &    380,    111  & Phobos\\
 Lick Observatory 1892, 1894 &   0.0083 &   0.3244 &  -0.0739 &   0.4133 &    305,    127 & Deimos\\
 \hline
\citet{1897AJ.....17..145B}, \citet{1910AJ.....26...69B}                                                 
  &  -0.0480 &   0.4101 &  -0.2146 &   0.3935 &     63,     50 & Phobos\\
 Lick 1894, Yerkes 1909 &   0.3599 &   0.3210 &  -0.0338 &   0.4927 &     18,     12 & Deimos\\
 \hline
 \citet{1989AAS...77..209M}, Telescope:26-inch                             
  &  -0.2101 &   0.3323 &  -0.0299 &   0.3433 &     24,     42  & Phobos\\
 U.S. Naval Obs., Washington (since 1893) &   0.0996 &   0.5336 &   0.0035 &   0.4300 &     52,     53 & Deimos\\
 \hline
\citet{1895MNRAS..55..348N}                                        
 &  -0.1343 &   0.4268 &   0.0240 &   0.7057 &     30,      3 & Phobos\\
Cambridge 1894 &   0.0000 &   0.0000 &   0.0000 &   0.0000 &      0,      0 & Deimos\\
 \hline
\citet{1913AJ.....27..163H}
 &   0.0426 &   0.4480 &   0.0387 &   0.2039 &     76,     79  & Phobos\\
USNO 1909  &   0.0006 &   0.3957 &   0.0306 &   0.3637 &     85,     85  & Deimos\\
 \hline
 \citet{1925LicOB..12....4J}                                                 
  &   0.0388 &   0.3907 &   0.0561 &   0.3054 &     41,     32  & Phobos\\
Lick 1924 &   0.1918 &   0.3629 &  -0.0064 &   0.2670 &     27,     21 & Deimos\\
\hline
\citet{1923AJ.....35..113B}                                                
  &   0.0391 &   0.3168 &   0.1281 &   0.1972 &     37,     38  & Phobos\\
USNO 1911-1922&   0.0498 &   0.2226 &  -0.0176 &   0.2165 &     50,     50 & Deimos\\
\hline 
\end{tabular} 
\end{center} 
\label{tab:moysigSP} 
\end{table*}

\begin{table*} 
\caption{Mean ($\nu$) and standard deviation ($\sigma$) on right ascension and declination in seconds of 
degrees for each satellite. $N$ is the number of observations by satellite.}
\begin{center} 
\begin{tabular}{lrrrrrl} 
\hline 
\hline 
Observations&$\nu_{\alpha\textcolor{black}{\cos\delta}}$&$\sigma_{\alpha\textcolor{black}{\cos\delta}}$&$\nu_\delta\ \ \ $&$\sigma_\delta\ \ \ $&$N\ \ $&satellite\\
& ('')$\ \ \  $& ('')$\ \ \  $& ('')$\ \ \  $& ('')$\ \ \  $&  & \\
\hline 
\citet{1989AAS...77..209M}, Hall-Newcomb-Harkness                                    
 &            -0.6395 &             0.0000 &             0.5384 &             0.0000 &      1,      1 & Phobos\\
 U.S. Naval Obs., Washington (before 1893) &            -0.9176 &             0.4875 &             1.3264 &             0.1682 &      4,      4  & Deimos\\
\hline
\citet{1880Obs.....3..270Y}                                                
 &            -1.0632 &             0.0000 &             0.0000 &             0.0000 &      1,      0& Phobos\\
Princeton 1879 &             0.0000 &             0.0000 &             0.0000 &             0.0000 &      0,      0 & Deimos\\
 \hline
\citet{1989AAS...77..209M}, U.S. Naval Obs., Washington (since 1893)                    
 &             0.0000 &             0.0000 &             0.0000 &             0.0000 &      0,      0 & Phobos\\
 USNO 1894 &            -1.0542 &             0.0000 &             0.3752 &             0.0000 &      1,      1 & Deimos\\
 \hline
\citet{1895MNRAS..55..348N}
 &             0.0283 &             0.7718 &            -0.2615 &             0.7440 &     23,     22& Phobos\\
 Cambridge 1894&             0.0000 &             0.0000 &             0.0000 &             0.0000 &      0,      0 & Deimos\\
  \hline
 \citet{1909AN....183....7K}, \citet{1913MiPul...5..149K}                                            
  &             0.0442 &             0.3596 &             0.0654 &             0.4658 &     23,     24 & Phobos\\
  Pulkovo 1909&             0.0029 &             0.5176 &            -0.0445 &             0.4297 &     22,     23& Deimos\\
 \hline
\citet{1913AJ.....27..163H}
  &            -0.3370 &             0.3341 &             0.0571 &             0.3727 &     38,     38 & Phobos\\
 USNO 1909 &            -0.5577 &             0.4243 &             0.2008 &             0.2360 &     31,     31 & Deimos\\
 \hline 
 \citet{Biesbroeck1970}                                      
 &            -0.1590 &             0.1340 &            -0.0217 &             0.0974 &     22,     22 & Phobos\\
Mc Donald 1956&             0.1590 &             0.1340 &             0.0217 &             0.0974 &     22,     22 & Deimos\\
 \hline
 \citet{2015AA...582A..36R}, 61-inch                                        
 &             0.0022 &             0.0532 &             0.0021 &             0.0514 &    216,    216 & Phobos\\
 USNO (Flagstaff) 1967-1986&             0.0018 &             0.0502 &            -0.0009 &             0.0534 &    254,    254& Deimos\\
 \hline
 \citet{2015AA...582A..36R},  26-inch                                        
  &             0.0002 &             0.0383 &             0.0024 &             0.0379 &    424,    424  & Phobos\\
 USNO 1971-1997&            -0.0006 &             0.0345 &            -0.0019 &             0.0343 &    450,    450 & Deimos\\
 \hline
\citet{1976IzPul.194..127K}                                             
  &            -0.0135 &             0.2281 &            -0.0280 &             0.2048 &     17,     17 & Phobos\\
 Pulkovo 1973 &             0.0135 &             0.2281 &             0.0280 &             0.2048 &     17,     17 & Deimos\\
 \hline
 \citet{1992Kudrya}                                              
  &             0.0097 &             0.0750 &            -0.0150 &             0.0901 &    660,    660 & Phobos\\
 Shokin Majdanak 1988 &            -0.0035 &             0.0753 &             0.0022 &             0.0718 &    639,    639 & Deimos\\
 \hline
 \citet{1991IzPul.207...37B}, inter-satellite only                                         
 &             0.0217 &             0.3642 &            -0.0520 &             0.2276 &     50,     50 & Phobos\\
Pulkovo 1988  &             0.0297 &             0.1824 &            -0.0171 &             0.2226 &     29,     29 & Deimos\\
 \hline
 \citet{1992AAS...96..485C}                                    
 &            -0.0215 &             0.0544 &            -0.0012 &             0.0639 &    813,    813 & Phobos\\
 Pic du Midi 1988 &             0.0215 &             0.0544 &             0.0012 &             0.0639 &    813,    813 & Deimos\\
 \hline
\citet{1989MNRAS.237P..15J}                                                
 &            -0.0043 &             0.1191 &             0.0506 &             0.1150 &    154,    154 & Phobos\\
 La Palma 1988&            -0.0574 &             0.1015 &            -0.0293 &             0.0984 &     78,     78 & Deimos\\
 \hline 
 Table Mountain Observatory (R.A.Jacobson, priv. com.)                                
 &             0.0244 &             0.0443 &             0.0112 &             0.1070 &      6,      6 & Phobos\\
 Table Mountain 2003&            -0.1105 &             0.1334 &            -0.0355 &             0.0961 &      9,      9  & Deimos\\
 \hline
  Pascu (priv. com.), B filter                                                  
 &             0.0116 &             0.0240 &             0.0104 &             0.0197 &     76,     76 & Phobos\\
 USNO (Flagstaff) 2003&            -0.0116 &             0.0240 &            -0.0104 &             0.0197 &     76,     76 & Deimos\\
 \hline
 Pascu (priv. com.), V filter                                                
 &             0.0025 &             0.0324 &             0.0105 &             0.0339 &     75,     75& Phobos\\
 USNO (Flagstaff) 2003&            -0.0025 &             0.0324 &            -0.0105 &             0.0339 &     75,     75  & Deimos\\
 \hline
 Pascu (priv. com.), R filter                                                     
 &            -0.0006 &             0.0267 &             0.0218 &             0.0363 &     56,     56 & Phobos\\
 USNO (Flagstaff) 2003&             0.0006 &             0.0267 &            -0.0218 &             0.0363 &     56,     56 & Deimos\\
\hline 
\end{tabular} 
\end{center} 
\label{tab:moysigRADEC} 
\end{table*}

\begin{table*} 
\caption{Mean ($\nu$) and standard deviation ($\sigma$) on right ascension and declination 
for each satellite. Both angles are multiplied by the distance spacecraft-moon to obtain kilometers. $N$ is the number of observations by satellite.
In the \citet{2015AA...580A..28P} publication positions of Phobos have been determined using control point (CP) and/or limb point (LF) measurements. The former are based on the satellite's control network, a set of identifiable surface features well-distributed over the body's surface and defining its reference system. Recent MEX SRC measurements have been distinguished into observations made during linear and non-linear pointing variations. If the observations' pre-fit residuals deviated by more than three sigma from the mean value, they have been additionally categorised as an outlier. In case of only a few outliers it could not be clarified whether they have been related to non-linear variations in pointing or not. Some measurements are based on fits of the shape model-derived limb to only very short limb point arcs in the image (see also subsection \ref{sec:close encounters}). }
\begin{center} 
\begin{tabular}{lrrrrrl} 
\hline 
\hline 
Observations&$\nu_{\alpha\textcolor{black}{\cos\delta}}$&$\sigma_{\alpha\textcolor{black}{\cos\delta}}$&$\nu_\delta\ \ \ $&$\sigma_\delta\ \ \ $&$N\ \ $&satellite\\
& (km)$\ \ \  $& (km)$\ \ \  $& (km)$\ \ \  $& (km)$\ \ \  $&  & \\
\hline
 Mariner 9         \citep{1989AA...216..284D}                                           
 &            -7.2594 &             7.3718 &            -6.2897 &             8.1181 &     48,     48 & Phobos\\
  &            -0.7282 &             4.1653 &            -3.4240 &             4.3704 &     14,     14  & Deimos\\
\hline
 Viking 1      \citep{1988AA...201..169D}                                        
 &            -0.4629 &             9.8178 &            -0.1971 &             8.7993 &    132,    132  & Phobos\\
 &             0.7174 &             3.4617 &            -1.9736 &             4.9500 &     19,     19   & Deimos\\
 \hline
 Viking 2     \citep{1988AA...201..169D}                                                
  &             2.6933 &             7.7604 &            -5.1783 &             8.0789 &     32,     32 & Phobos\\
 &            -1.4065 &             3.3602 &            -0.7202 &            11.9342 &     80,     80 & Deimos\\
 \hline
 Phobos 2      \citep{1991AA...244..236K}     
 &            -0.3396 &             0.8558 &            -0.2277 &             0.5194 &     37,     37 & Phobos\\
  &             3.9608 &            15.3037 &            12.0239 &            11.2175 &      8,      8 & Deimos\\
\hline 
 MEX  (\citep{2008AA...488..361W}; priv. com.)                                 
 &            -0.0867 &             0.4485 &             0.0548 &             0.4973 &    135,    135 & Phobos\\
 &             0.0000 &             0.0000 &             0.0000 &             0.0000 &      0,      0  & Deimos\\
 \hline
 MEX    \citep{2012AA...545A.144P}                                             
 &             0.0000 &             0.0000 &             0.0000 &             0.0000 &      0,      0 & Phobos\\
  &             0.0395 &             1.1084 &            -0.3801 &             1.0322 &    136,    136  & Deimos\\
 \hline
 MEX (CP)    \citep{2015AA...580A..28P}                                 
 &            -0.3113 &             0.5400 &             0.0383 &             0.9030 &    130,    130& Phobos\\
 &             0.0000 &             0.0000 &             0.0000 &             0.0000 &      0,      0 & Deimos\\
 \hline
 MEX (LF)       \citep{2015AA...580A..28P}                               
  &            -0.1153 &             0.4126 &             0.2043 &             0.9995 &     27,     27 & Phobos\\
 &             0.0000 &             0.0000 &             0.0000 &             0.0000 &      0,      0  & Deimos\\
 \hline
 MEX-flyby-linear-outlier       (Pasewaldt et al., in prep)                           
 &            -0.3070 &             0.0529 &             0.4736 &             0.0411 &      3,      3  & Phobos\\
 &             0.0000 &             0.0000 &             0.0000 &             0.0000 &      0,      0  & Deimos\\
 \hline
 MEX-flyby-linear            (Pasewaldt et al., in prep)                                  
  &            -0.0872 &             0.2683 &             0.0488 &             0.3004 &     64,     64 & Phobos\\
 &             0.0000 &             0.0000 &             0.0000 &             0.0000 &      0,      0  & Deimos\\
 \hline
 MEX-flyby-non-linear          (Pasewaldt et al., in prep)                                 
 &             0.2143 &             0.4333 &             0.1653 &             0.3507 &     22,     22  & Phobos\\
 &             0.0000 &             0.0000 &             0.0000 &             0.0000 &      0,      0  & Deimos\\
 \hline
 MEX-flyby-non-linear-outlier           (Pasewaldt et al., in prep)                        
 &            -0.0224 &             0.2974 &             0.2931 &             0.1644 &      8,      8 & Phobos\\
 &             0.0000 &             0.0000 &             0.0000 &             0.0000 &      0,      0  & Deimos\\
 \hline
 MEX-flyby-shortlimb     (Pasewaldt et al., in prep)                                       
  &            -0.4813 &             0.2241 &             0.3064 &             0.4526 &      6,      6 & Phobos\\
 &             0.0000 &             0.0000 &             0.0000 &             0.0000 &      0,      0  & Deimos\\
 \hline
 MEX-flyby-outlier           (Pasewaldt et al., in prep)                                   
 &            -0.8951 &             0.0746 &            -0.5160 &             0.1332 &      3,      3& Phobos\\
 &             0.0000 &             0.0000 &             0.0000 &             0.0000 &      0,      0   & Deimos\\
  \hline
 MEX-Dei-Saturn-Ziese                                        
 &             0.0000 &             0.0000 &             0.0000 &             0.0000 &      0,      0 & Phobos\\
 &            -0.2029 &             0.1710 &            -0.2820 &             0.0931 &     64,     64 & Deimos\\
 \hline
 MEX-Pho-Saturn-Ziese                                        
 &             0.1904 &             0.2440 &             0.1121 &             0.1287 &    295,    295 & Phobos\\
 &             0.0000 &             0.0000 &             0.0000 &             0.0000 &      0,      0 & Deimos\\
 \hline
 MEX-Dei-Jup-Ziese                                           
 &             0.0000 &             0.0000 &             0.0000 &             0.0000 &      0,      0 & Phobos\\
 &            -1.0983 &             0.0782 &             0.8743 &             0.0184 &     14,     14 & Deimos\\
  \hline
 MEX-Pho-Jup-Ziese                                           
 &             0.2920 &             0.0269 &            -0.0773 &             0.0319 &     50,     50 & Phobos\\
 &             0.0000 &             0.0000 &             0.0000 &             0.0000 &      0,      0 & Deimos\\
\hline 
\end{tabular} 
\end{center} 
\label{tab:moysig} 
\end{table*}

\begin{table*} 
\caption{Mean ($\nu$) and standard deviation ($\sigma$) on sample and line in pixel \textcolor{black}{and kilometer} for each satellite. $N$ is the number of observations by satellite.
MRO (single) gathers data where only one moon was observable at a time.}
\begin{center} 
\begin{tabular}{lrrrrrrrrrl} 
\hline 
\hline 
Observations&$\nu_{\mathrm{sample}}$&&$\sigma_{\mathrm{sample}}$&&$\nu_{\mathrm{line}}$&&$\sigma_{\mathrm{line}}$&&$N\ \ $&satellite\\
&(pix)\ \ \ $$& $(km)\ \ \ \ $& (pix)\ \ \ $$& $(km)\ \ \ \ $& (pix)\ \ \ $$& $(km)\ \ \ \ $& (pix)\ \ \ $$& $(km)\ \ \ \ $&  & \\
 \hline
 MRO (single)                                                
  &             0.0000 &    0.0000 &         0.0000 &      0.0000 &       0.0000 &       0.0000 &      0.0000 &   0.0000 &   0,      0 & Phobos\\
  &             0.0031 &   -0.0416&          0.0722 &   3.6539&          0.0081 &       0.8250&      0.0614 &  3.500&  376,    376 & Deimos\\
 \hline
 MRO                                                      
  &             0.0062 &   -0.0607&       0.0466 &    3.5902&      -0.0140 &     0.1301&    0.0723 &  4.3938&   103,    103 & Phobos\\
  &            -0.0062 &    0.0535&       0.0466 &    3.5921&       0.0140 &    -0.1201&    0.0723 &  4.4041&   103,    103 & Deimos\\
 \hline
\end{tabular} 
\end{center} 
\label{tab:moysig2} 
\end{table*}

\begin{table*} 
\caption{\textcolor{black}{Initial conditions and related uncertainties of Phobos and Deimos in the ICRF after fit at initial epoch J2000 (Julian day 2451545.0). Units are $km$ and $km/sec$. All digits have been kept for reproducibility of our results.}}
\begin{center} 
\begin{tabular}{lrrr} 
\hline 
\hline 
Moon&&&\\
\hline
Phobos &&&\\
$x, y, z$ & -1989.71893421683  $\pm$  0.093  &   -8743.02171778182   $\pm$  0.025 &   -3181.65223492620    $\pm$  0.065 \\
$vx, vy, vz$ &  1.84320501057858  $\pm$  3.8E-6 &    -4.312869534231428E-2  $\pm$ 2.6E-6 & -1.01836853901114  $\pm$ 2.1E-6  \\
Deimos &&&\\
$x, y, z$ &    10366.3782389459   $\pm$ 0.46   &  -15747.7651557215   $\pm$ 1.42 &    -13945.1083381706    $\pm$ 0.69  \\
$vx, vy, vz$  & 1.04085057050574   $\pm$  1.09E-5 &   0.843538311722695  $\pm$ 1.09E-5  &    -0.178915197575563  $\pm$  1.10E-5 \\
\hline
\end{tabular} 
\end{center} 
\label{tab:CI}
\end{table*}

\begin{table*} 
\caption{Estimated forced libration in longitude of Phobos. \citet{1974Icar...23..290D}, \citet{1989AA...216..284D},  \citet{2010EPSL.294..541W} and \citet{2018JGeod..92..963B} used spacecraft imaging to solve directly for Phobos' physical libration. \citet{1990AA...233..235B} and \citet{2014PSS..102...51W} computed Phobos' libration from the observed shape of Phobos assuming a homogeneous interior. Here, we follow \citet{2010AJ....139..668J} and determine Phobos' physical libration from its effect on Phobos' orbit.}
\begin{center} 
\begin{tabular}{lr} 
\hline 
\hline 
Reference & $A$ (deg)\\
\hline 
\citet{1974Icar...23..290D} & 3. \\
\hline
\citet{1989AA...216..284D} & $0.81 \pm 0.5$ \\
\hline
\citet{1990AA...233..235B} &$ 1.19 $\\
\hline
\citet{2010AJ....139..668J} &$ 1.03 \pm 0.22$ \\
\hline
\citet{2010EPSL.294..541W}& $1.20 \pm 0.14$ \\
\hline
\citet{2014PSS..102...51W} & $ 1.14 $ \\
\hline
\citet{2018JGeod..92..963B} & $ 1.14 \pm 0.03 $ \\
\hline
present work ($1\sigma$)&$ 1.09 \pm 0.01 $\\
\hline 
\end{tabular} 
\end{center} 
\label{tab:libration} 
\end{table*}

\section{Precision versus accuracy}\label{sec:5}

While fitting our model to astrometric data, we discovered a disagreement between MRO\footnote{MRO astrometric data are available at: 
http://ssd.jpl.nasa.gov/dat/sat/psf\_mro.txt.} and MEX data. In particular, while both sets did provide small residuals independently, they appeared to be strongly deteriorated when considered together for Deimos. Since both data sets roughly cover the same years, an error on the modeling sounds unlikely. Unfortunately, we could not use Phobos data to get a deeper understanding of the possible bias in these sets since MRO data of Phobos are of much less quality. This bias was simultaneously observed at JPL and interpreted as a bias of $55.''775$ and $-7.''321$ in right ascension and declination of MEX data, respectively (Robert Jacobson, private communication). Since this bias appeared roughly constant over years, MEX observations were supposed to be offset. However, due to MEX' specific orbit, most MEX observations were performed in similar geometrical condition. As a consequence, a constant bias on the predicted position of Deimos may provide similar astrometric residuals.

A closer look at the MRO data showed that the errors of a few kilometers on Deimos observations were related to a difference lower than 0.2 pixel, only. Such a small difference is related to the scale of the image since Deimos was far away when observed for astrometric purposes by the MRO spacecraft. In particular, the Martian moons are known to have a complex shape, and the fit of their center of figure (assumed to be approximately equal to their center of mass) may be inaccurate when not properly resolved. 
Last but not least, using independent measurements, \citet{2018A&A...614A..15Z} showed that our former solution NOE-4-2015-b \textcolor{black}{(which did not use MRO data)} was in much better agreement for Deimos than mar097.
Hence, we privileged here again the MEX data and \textcolor{black}{removed} the MRO data during the fitting procedure.

\section{Comparison of ephemerides and extrapolation precision}\label{sec:6}

\begin{figure*} 
\begin{center} 
\begin{tabular}{ll} 
\hspace*{-0.35cm}\includegraphics[width=9.5cm,angle=0]{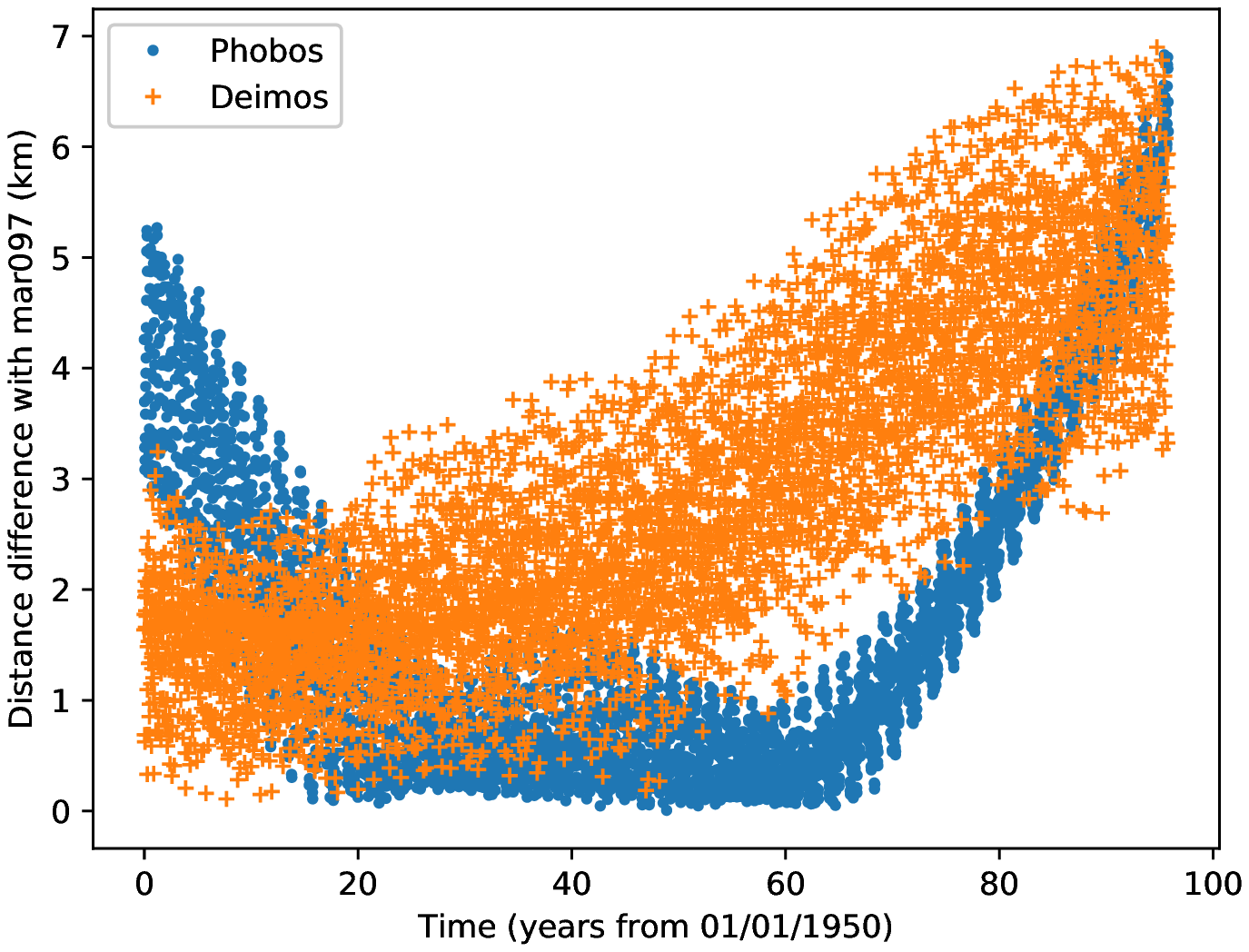} & \includegraphics[width=9.5cm,angle=0]{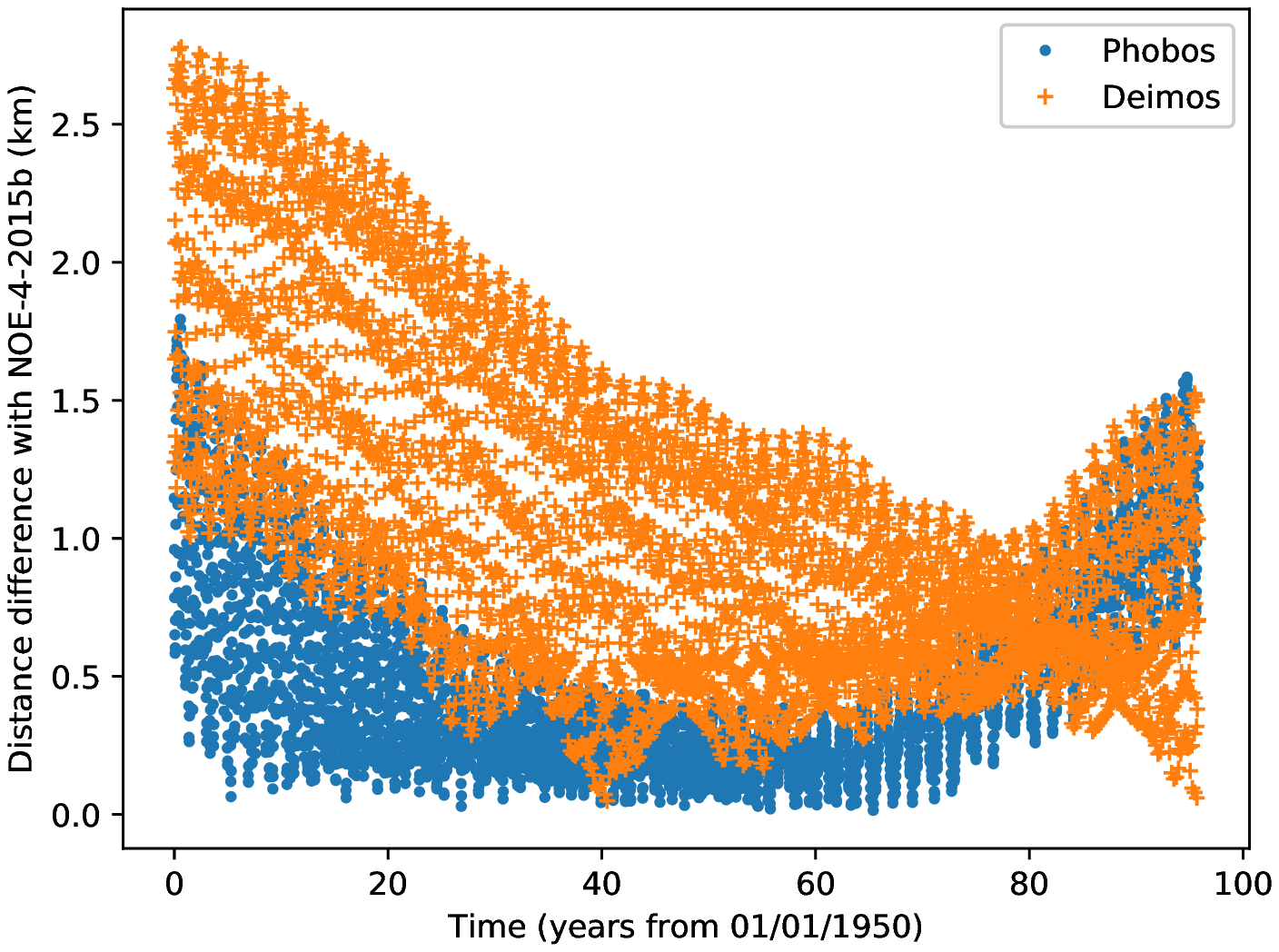}\\
\end{tabular} 
\caption{Differences in distance in 3-dimensional space between our new ephemerides (NOE-4-2020) and: i) the JPL ephemerides mar097 (left); ii) our former ephemerides NOE-4-2015b (right). }\label{fig:comparision-ephemeris}
\end{center} 
\end{figure*}

We compared our ephemerides with our former solution NOE-4-2015-b as well as the ephemerides mar097 developed at JPL \citep{2010AJ....139..668J}. \textcolor{black}{In particular, NOE-4-2015-b solution used a pretty similar dynamical model and observation sets to the present work. However, it do not benefit from the new MEX observations of \citet{2018A&A...614A..15Z}, as well as the unpublished ones used here.}
Figure \ref{fig:comparision-ephemeris} shows differences for both moons in distance in 3-dimensional space. In the case of Phobos the differences at the time close to 2010 are particularly small (at the level of few kilometers) as a consequence of MEX data that have already been available in 2010. Then, differences increase with time due to small modeling/weight differences during the fitting procedure. 
The new MEX data used in this study shall foster our confidence in the current ephemerides. On the other hand, in the case of Deimos differences increase linearly over time, as a consequence of the different treatment of the MRO data. As already said, such difference was already pointed out by \citet{2018A&A...614A..15Z}.

\begin{figure*} 
\begin{center} 
\begin{tabular}{ll} 
\hspace*{-0.35cm}\includegraphics[width=9.5cm,angle=0]{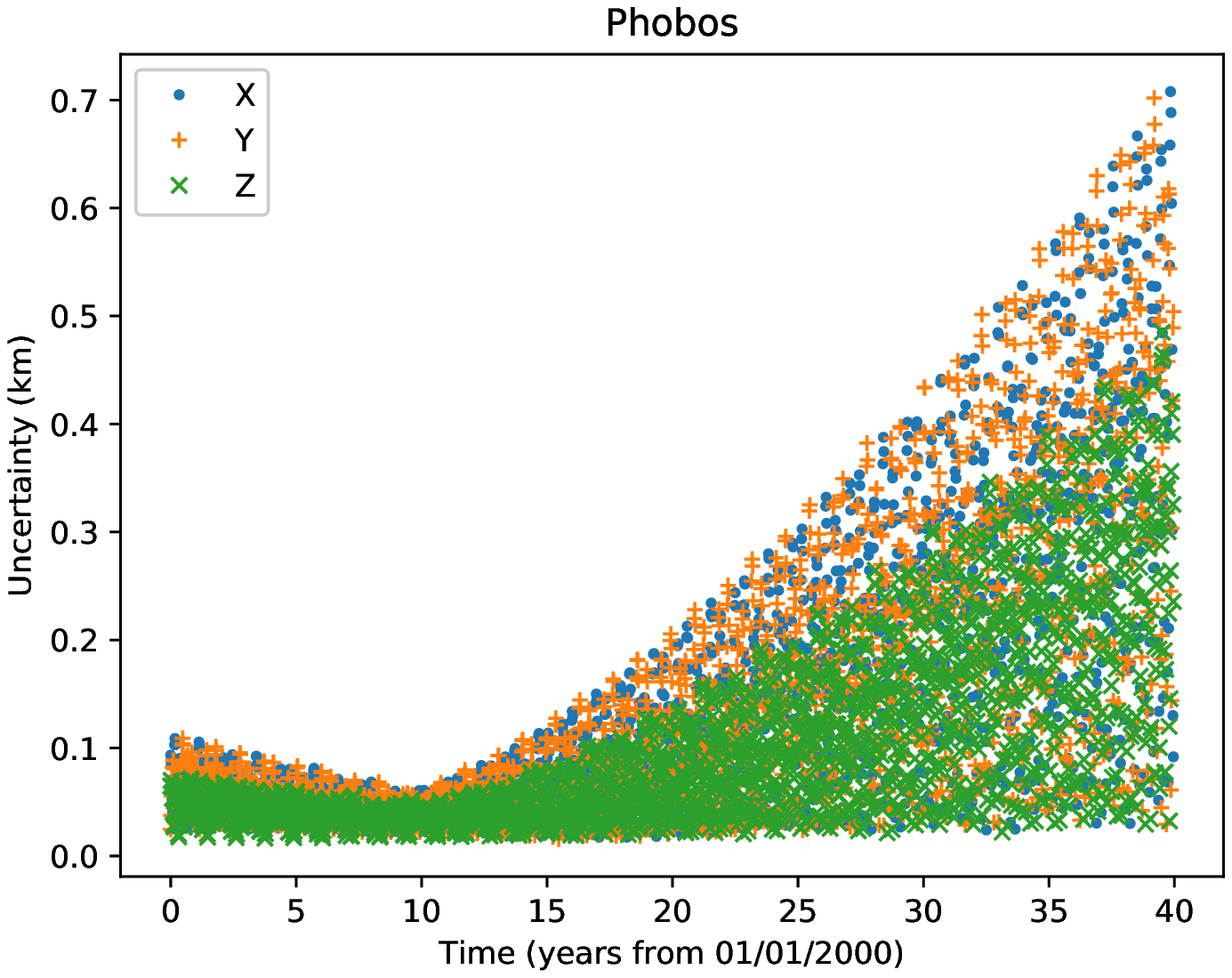} & \includegraphics[width=9.5cm,angle=0]{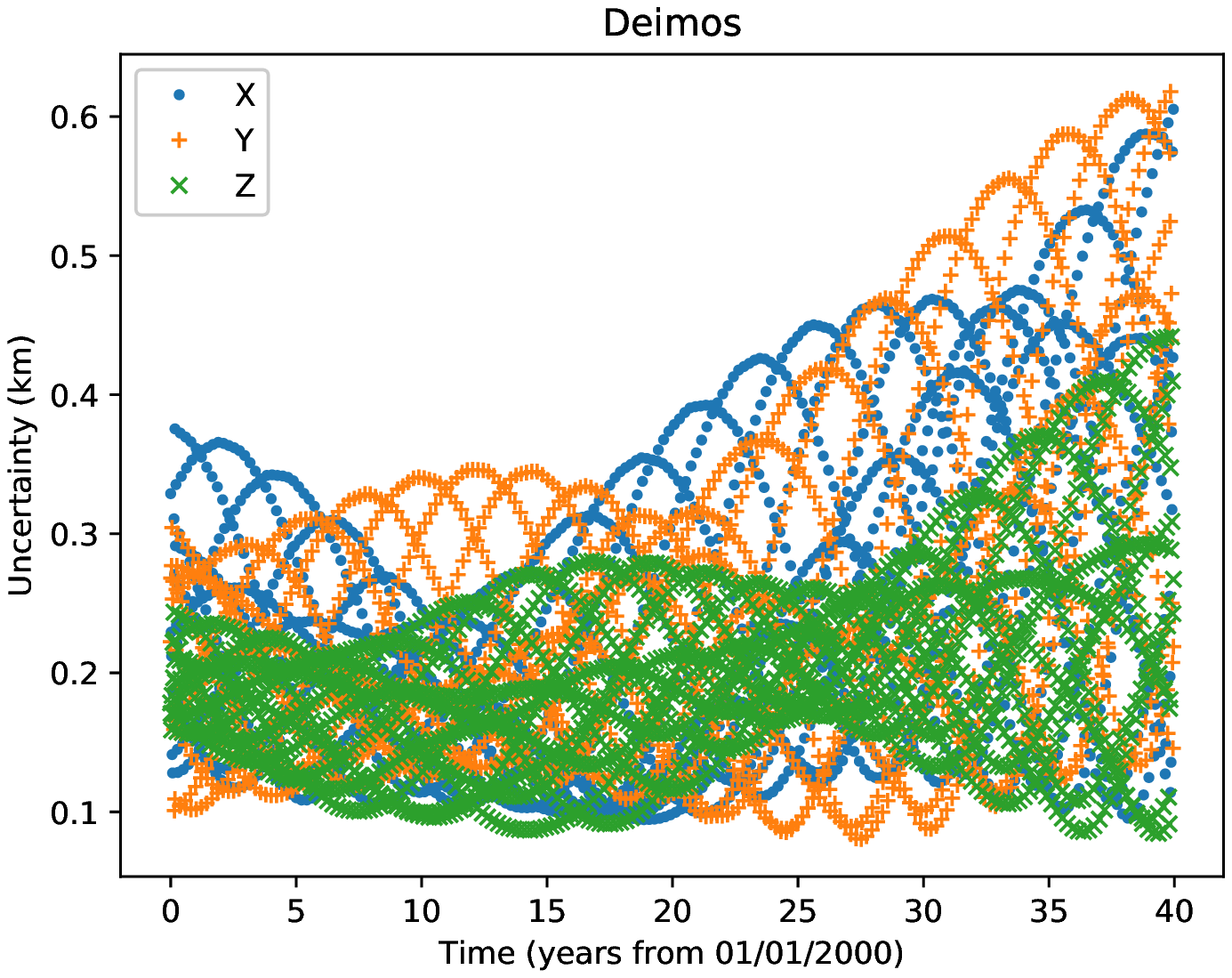}
\end{tabular} 
\caption{Ephemerides 1$\sigma$ uncertainty.}\label{fig:1-sigma-uncertainty}
\end{center} 
\end{figure*}

Figure \ref{fig:1-sigma-uncertainty} shows the formal uncertainty of our new ephemeris of Phobos and Deimos. Phobos' ephemeris uncertainty is typically few hundred of meters for the present time. Due to its larger distance to MEX, Deimos ephemeris is less well constrained, but expected to be precise within several hundreds of meters. The real accuracy of our new ephemerides will have to be confirmed by confrontation with independent observational means.

\section{Conclusions}

We developed new ephemerides of Phobos and Deimos, fitted to the largest time span of observations, including recent unpublished astrometric data obtained by evaluating images taken by the HRSC aboard Mars Express. During the fitting process, we could solve for the physical libration of Phobos, found to be equal to $1.09 \pm 0.01$ degrees. Considering the very low uncertainty, this may suggest to consider higher order harmonics with an improved rotation model in the future. We confirmed an inconsistency between recent Mars Reconnaissance Orbiter and Mars Express  data on Deimos observations. Since Mars Reconnaissance Orbiter's observations were performed from a large distance we imputed the inconsistency to an erroneous center of figure determination. Our ephemerides NOE-4-2020 are available in SPICE format directly from ftp://ftp.imcce.fr/pub/ephem/satel/NOE/MARS/2020/.

\begin{acknowledgements}
The authors are indebted to O.\added[id=ra]{ }Witasse, R.\added[id=ra]{ }Jacobson and S.\added[id=ra]{ }Le Maistre for fruitful discussions.
This work has been supported by the European Community's Seventh Framework Program (FP7/2007-2013) under grant agreement 263466 for the FP7-ESPaCE project. 
The authors wish to thank the HRSC Experiment team at DLR, Institute of Planetary Research, Berlin, and at
Freie Universit\"at Berlin, the HRSC Science Team, as well as the Mars Express Project teams at
ESTEC, ESOC, and ESAC for their successful planning, acquisition, and release of image data to the community.
\end{acknowledgements}

\bibliographystyle{aa} 


\end{document}